\pdfoutput=1
\documentclass[12pt,a4paper]{article}

\usepackage{ifthen} 
\usepackage{multirow}
\usepackage{subfig} 

\usepackage{longtable}
\usepackage{rotating}

\newboolean{pdflatex}
\setboolean{pdflatex}{true} 

\newboolean{articletitles}
\setboolean{articletitles}{true} 

\newboolean{uprightparticles}
\setboolean{uprightparticles}{false} 

\newboolean{inbibliography}
\setboolean{inbibliography}{false} 

\def\delpar{\delta_\parallel}
\def\delperp{\delta_\perp}

\def\grpsuthree {{\ensuremath{\mathrm{SU}(3)}}\xspace}
\newcommand{\CP}{\ensuremath{{C\!P}}\xspace}

\newcommand{\ACPL}{\ensuremath{A^{C\!P}_0}\xspace}
\newcommand{\ACPpa}{\ensuremath{A^{C\!P}_{\|}}\xspace}
\newcommand{\ACPpe}{\ensuremath{A^{C\!P}_{\perp}}\xspace}
\newcommand{\ACPS}{\ensuremath{A^{C\!P}_{\rm S}}\xspace}

\newcommand{\fL}{\ensuremath{f_0}\xspace}
\newcommand{\fpa}{\ensuremath{f_{\|}}\xspace}

\def\thetamu{\ensuremath{\theta_{\mu}}\xspace}
\def\thetaK{\ensuremath{\theta_{K}}\xspace}
\def\phihel{\ensuremath{\varphi_{h}}\xspace}

\newcommand{\splot}  {\mbox{$sP\!lot$}\xspace}
\newcommand{\sweights} {\mbox{$sW\!eights$}\xspace}

\def\photos     {\mbox{\textsc{Photos}}\xspace}
\def\pythia     {\mbox{\textsc{Pythia}}\xspace}
\def\evtgen     {\mbox{\textsc{EvtGen}}\xspace}
\def\geant      {\mbox{\textsc{Geant4}}\xspace}

\def\ptot       {\mbox{$p$}\xspace}
\def\pt         {\mbox{$p_{\rm T}$}\xspace}

 \def\PJ      {\ensuremath{J}\xspace}

 \def\Pb      {\ensuremath{b}\xspace}                 
 \def\Pc      {\ensuremath{c}\xspace}

 \def\Pi      {\ensuremath{i}\xspace}

\def\cquark    {\ensuremath{\Pc}\xspace}

\def\bquark    {\ensuremath{\Pb}\xspace}

\newcommand{\rad}{\ensuremath{\rm \,rad}\xspace}

\def\cosTmu  {\ensuremath{\cos(\theta_{\mu})}\xspace}

\def\lhcb {\mbox{LHCb}\xspace}
\def\mum  {\ensuremath{\rm \mu m}\xspace}

\def\CP                {\ensuremath{C\!P}\xspace}

\def\Ppsi {\ensuremath{\psi}\xspace}
\def\jpsi     {{\ensuremath{{\PJ\mskip -3mu/\mskip -2mu\Ppsi\mskip 2mu}}}\xspace}
\def\Jpsi     {{\ensuremath{\jpsi}}\xspace}

\newcommand{\mkpi}{\ensuremath{m_{\Km\pip}}\xspace}

\def\Bbar    {\kern 0.18em\overline{\kern -0.18em B}{}\xspace}

\newcommand{\Bs}{\ensuremath{B^0_s}\xspace}

\def\Bsb     {\ensuremath{\Bbar^0_s}\xspace}

\newcommand{\Bd}{\ensuremath{B^0}\xspace}
\def\Bdb     {\ensuremath{\Bbar^0}\xspace}
\newcommand{\Bu}{\ensuremath{B^+}\xspace}
\newcommand{\Bub}{\ensuremath{B^-}\xspace}

\newcommand{\Dstpm}{\ensuremath{D^{*\pm}}\xspace}

\mathchardef\PLambda="7103
\newcommand{\Lambdab}{\ensuremath{\PLambda^0_b}\xspace}
\newcommand{\Lb}{\Lambdab}

\newcommand{\Kst}{\ensuremath{K^{*0}}\xspace}
\newcommand{\BJpsiX}{\ensuremath{B_{q}\to \jpsi X}\xspace}
\newcommand{\BJpsihh}{\ensuremath{B\to \jpsi h^+h^-}\xspace}
\newcommand{\bJpsihh}{\ensuremath{b{\text{-hadron}}~{\text{to}}~\jpsi h^+h^-}\xspace}

\newcommand{\BdJpsiKst}{\ensuremath{B^0\to \jpsi K^{*0}}\xspace}
\newcommand{\BdJpsiKpi}{\ensuremath{B^0\to \jpsi \Kp \pim}\xspace}

\newcommand{\BsJpsiKpi}{\ensuremath{B^0_s\to \jpsi \Km \pip}\xspace}

\newcommand{\BsJpsiKst}{\ensuremath{B^0_s\to \jpsi \Kstarzb}\xspace}
\newcommand{\BJpsiKst}{\ensuremath{B^0_{(s)}\to \jpsi K^{*0}(\Kstarzb)}\xspace}

\newcommand{\BdJpsipipi}{\ensuremath{B^0\to \jpsi \pi^+\pi^-}\xspace}
\newcommand{\BsJpsipipi}{\ensuremath{B^0_s\to \jpsi \pi^+\pi^-}\xspace}

\newcommand{\BsJpsiKK}{\ensuremath{B^0_s\to \jpsi K^+K^-}\xspace}

\newcommand{\LbJpsipK}{\ensuremath{\Lambdab\to\Jpsi p K^-}\xspace}
\newcommand{\LbJpsippi}{\ensuremath{\Lambdab\to\Jpsi p \pi^-}\xspace}
\newcommand{\LbJpsiph}{\ensuremath{\Lambdab\to\Jpsi p h^-}\xspace}
\newcommand{\BsJpsiPhi}{\ensuremath{B^0_s\to \jpsi \phi}\xspace}

\newcommand{\BdJKst}{\BdJpsiKst}

\newcommand{\BsJKst}{\BsJpsiKst}
\newcommand{\BsJphi}{\BsJpsiPhi}

\newcommand{\phis}{\ensuremath{\phi_s}\xspace}
\newcommand{\invfb}{\ensuremath{\mbox{\,fb}^{-1}}\xspace}

\def\Kbar  {\kern 0.2em\overline{\kern -0.2em K}{}\xspace}
\def\Kstarzb {\ensuremath{\Kbar^{*0}}\xspace}

\newcommand{\BRof}[1]{\ensuremath{{\cal B}(#1)}\xspace}

\newcommand{\BR}{\BRof}

\newcommand{\tev}{\ifthenelse{\boolean{inbibliography}}{\ensuremath{~T\kern -0.05em eV}\xspace}{\ensuremath{\mathrm{\,Te\kern -0.1em V}}}\xspace}
\newcommand{\gev}{\ensuremath{\mathrm{\,Ge\kern -0.1em V}}\xspace}
\newcommand{\mev}{\ensuremath{\mathrm{\,Me\kern -0.1em V}}\xspace}
\newcommand{\kev}{\ensuremath{\mathrm{\,ke\kern -0.1em V}}\xspace}
\newcommand{\ev}{\ensuremath{\mathrm{\,e\kern -0.1em V}}\xspace}
\newcommand{\gevc}{\ensuremath{{\mathrm{\,Ge\kern -0.1em V\!/}c}}\xspace}
\newcommand{\mevc}{\ensuremath{{\mathrm{\,Me\kern -0.1em V\!/}c}}\xspace}
\newcommand{\gevcc}{\ensuremath{{\mathrm{\,Ge\kern -0.1em V\!/}c^2}}\xspace}
\newcommand{\gevgevcccc}{\ensuremath{{\mathrm{\,Ge\kern -0.1em V^2\!/}c^4}}\xspace}
\newcommand{\mevcc}{\ensuremath{{\mathrm{\,Me\kern -0.1em V\!/}c^2}}\xspace}

\newcommand{\TeV}{\tev}

\newcommand{\MeVcc}{\mevcc}

\newcommand{\Km}{\ensuremath{K^-}\xspace}
\newcommand{\Kp}{\ensuremath{K^+}\xspace}
\newcommand{\KS  }{\ensuremath{K^0_{\mathrm{\scriptscriptstyle S}}}\xspace} 
\newcommand{\pim}{\ensuremath{\pi^-}\xspace}
\newcommand{\pip}{\ensuremath{\pi^+}\xspace}
\newcommand{\piz}{\ensuremath{\pi^0}\xspace}

\newcommand{\pdf}{\ensuremath{{\rm PDF}}\xspace}

\newcommand{\swave}{{\rm S--wave}\xspace}
\newcommand{\pwave}{{\rm P--wave}\xspace}
\newcommand{\dwave}{{\rm D--wave}\xspace}

\newcommand{\equref}[1]{Eq.~\ref{#1}}
\newcommand{\figref}[1]{Fig.~\ref{#1}}
\newcommand{\tabref}[1]{Table~\ref{#1}}
\newcommand{\appref}[1]{Appendix~\ref{#1}}

\newcommand{\secref}[1]{Sect.~\ref{#1}}

\usepackage{microtype}
\usepackage{lineno}  
\usepackage{xspace} 
\usepackage{caption} 

\usepackage{graphicx}  
\usepackage{color}
\usepackage{colortbl}
\graphicspath{{./figs/}} 

\usepackage{amsmath} 
\usepackage{amssymb}
\usepackage{amsfonts}
\usepackage{upgreek} 

\newcommand*\patchAmsMathEnvironmentForLineno[1]{%
\expandafter\let\csname old#1\expandafter\endcsname\csname #1\endcsname
\expandafter\let\csname oldend#1\expandafter\endcsname\csname
end#1\endcsname
 \renewenvironment{#1}%
   {\linenomath\csname old#1\endcsname}%
   {\csname oldend#1\endcsname\endlinenomath}%
}
\newcommand*\patchBothAmsMathEnvironmentsForLineno[1]{%
  \patchAmsMathEnvironmentForLineno{#1}%
  \patchAmsMathEnvironmentForLineno{#1*}%
}
\AtBeginDocument{%
\patchBothAmsMathEnvironmentsForLineno{equation}%
\patchBothAmsMathEnvironmentsForLineno{align}%
\patchBothAmsMathEnvironmentsForLineno{flalign}%
\patchBothAmsMathEnvironmentsForLineno{alignat}%
\patchBothAmsMathEnvironmentsForLineno{gather}%
\patchBothAmsMathEnvironmentsForLineno{multline}%
\patchBothAmsMathEnvironmentsForLineno{eqnarray}%
}

\usepackage{hyperref}    
\usepackage[all]{hypcap} 

\usepackage{cite} 
\usepackage{mciteplus}

\usepackage{longtable}

\makeatletter
\g@addto@macro\bfseries{\boldmath}
\makeatother

\begin{document}

\renewcommand{\thefootnote}{\fnsymbol{footnote}}
\setcounter{footnote}{1}

\begin{titlepage}
\pagenumbering{roman}

\vspace*{-1.5cm}
\centerline{\large EUROPEAN ORGANIZATION FOR NUCLEAR RESEARCH (CERN)}
\vspace*{1.5cm}
\hspace*{-0.5cm}
\begin{tabular*}{\linewidth}{lc@{\extracolsep{\fill}}r}
\ifthenelse{\boolean{pdflatex}}
{\vspace*{-2.7cm}\mbox{\!\!\!\includegraphics[width=.14\textwidth]{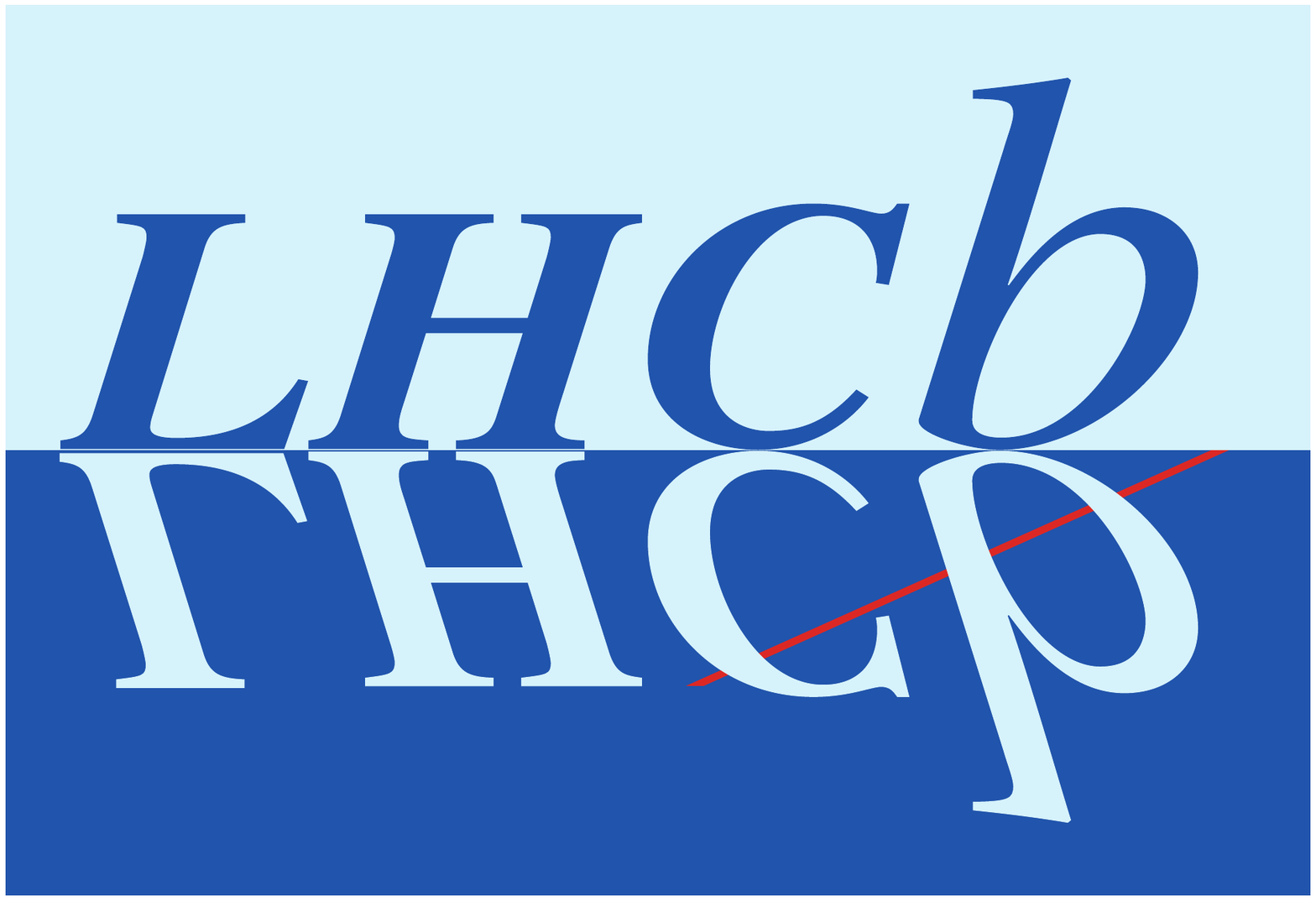}} & &}
{\vspace*{-1.2cm}\mbox{\!\!\!\includegraphics[width=.12\textwidth]{lhcb-logo.eps}} & &}
\\
 & & CERN-PH-EP-2015-224 \\  
 & & LHCb-PAPER-2015-034 \\  
 & & \today \\
 & & \\
\end{tabular*}

\vspace*{1.0cm}

{\bf\boldmath\huge
\begin{center}
 Measurement of \CP violation parameters and polarisation fractions in $B_s^0\to \jpsi \Kstarzb$ decays
\end{center}
}

\vspace*{1.0cm}

\begin{center}
The LHCb collaboration\footnote{Authors are listed at the end of this paper.}
\end{center}

\abstract{
\noindent
The first measurement of \CP{} asymmetries in the decay $B^0_s\to \jpsi \Kbar^{*}(892)^0$ and an updated measurement of its branching fraction and polarisation fractions are presented.
The results are obtained using data corresponding to an integrated luminosity of 3.0\:\invfb of proton--proton collisions recorded with the LHCb detector at centre-of-mass energies of 7 and 8\TeV.
Together with constraints from $\Bd\to \jpsi \rho^0$, the results are used to constrain additional contributions due to penguin diagrams in the \CP-violating phase $\phi_s$, measured through $\Bs$ decays to charmonium.
}

\vspace*{5.0cm}

\begin{center}
  Published in JHEP 
\end{center}

\vspace{\fill}

{\footnotesize 
\centerline{\copyright~CERN on behalf of the \lhcb collaboration, licence \href{http://creativecommons.org/licenses/by/4.0/}{CC-BY-4.0}.}}
\vspace*{2mm}

\end{titlepage}

\newpage
\setcounter{page}{2}
\mbox{~}

\cleardoublepage

\renewcommand{\thefootnote}{\arabic{footnote}}
\setcounter{footnote}{0}


\pagestyle{plain} 
\setcounter{page}{1}
\pagenumbering{arabic}

\section{Introduction}
\label{sec:intro}

\begin{figure}[t]
\includegraphics[width=0.49\textwidth]{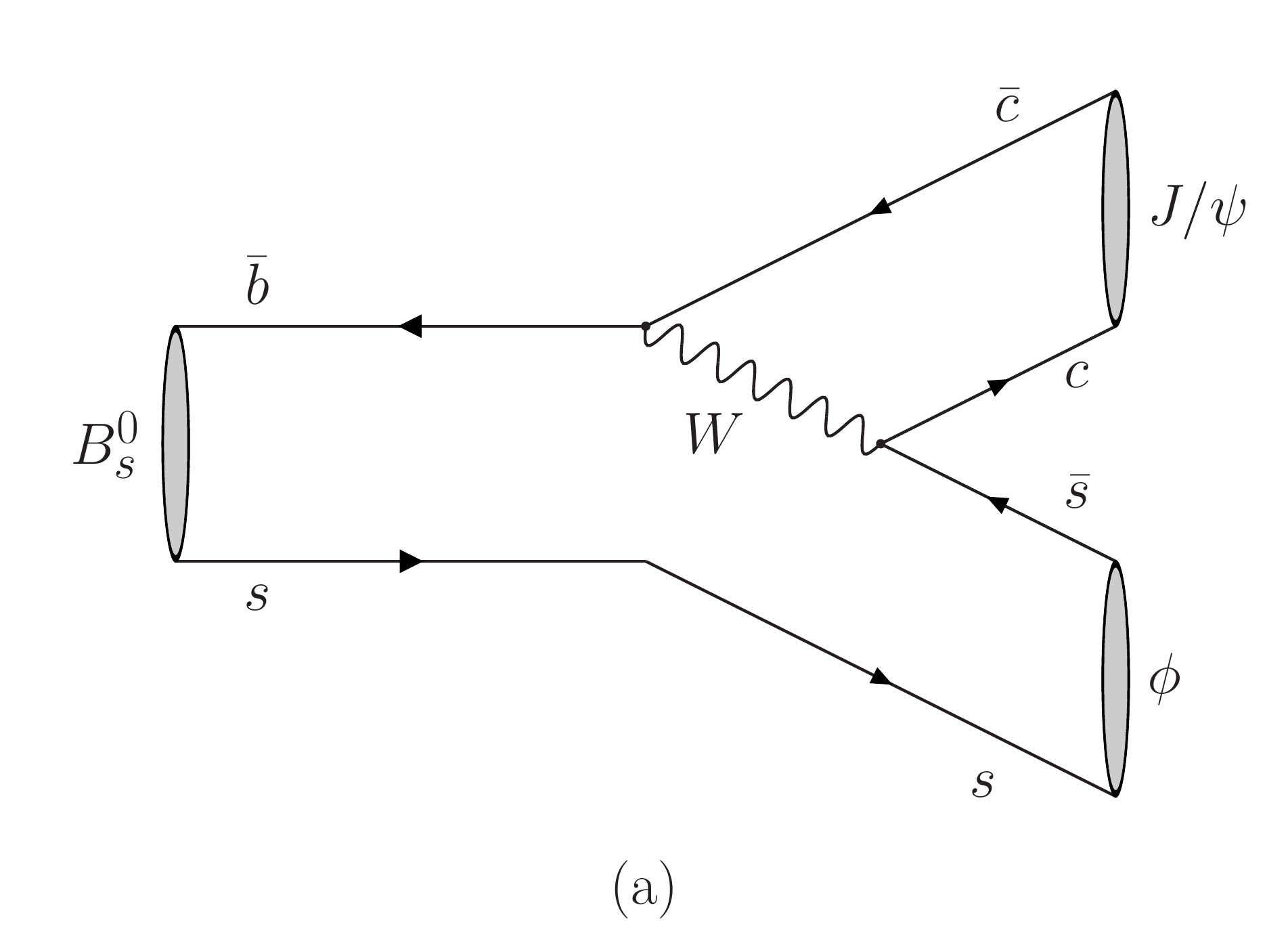}
\hfill
\includegraphics[width=0.49\textwidth]{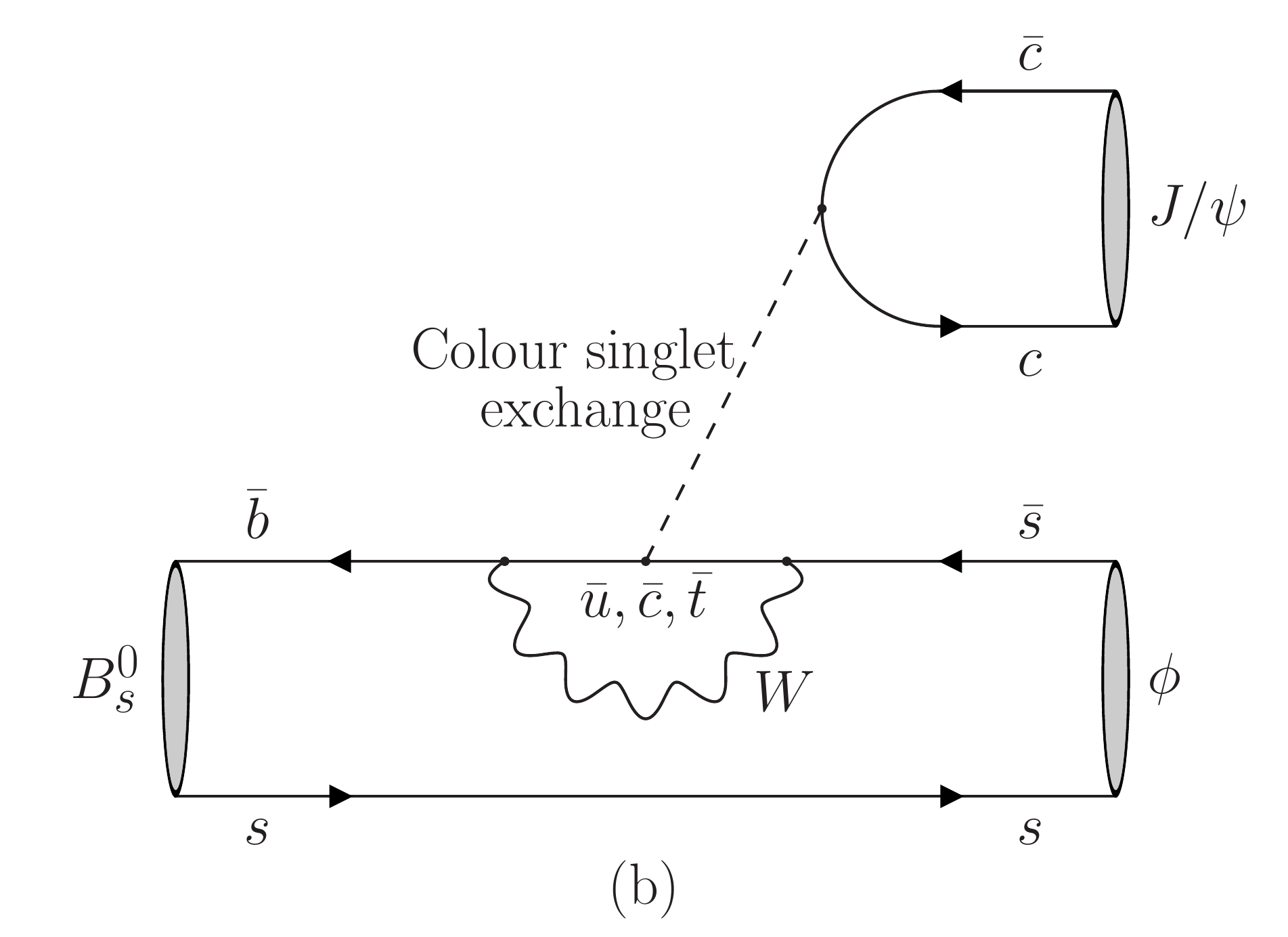} \\
\includegraphics[width=0.49\textwidth]{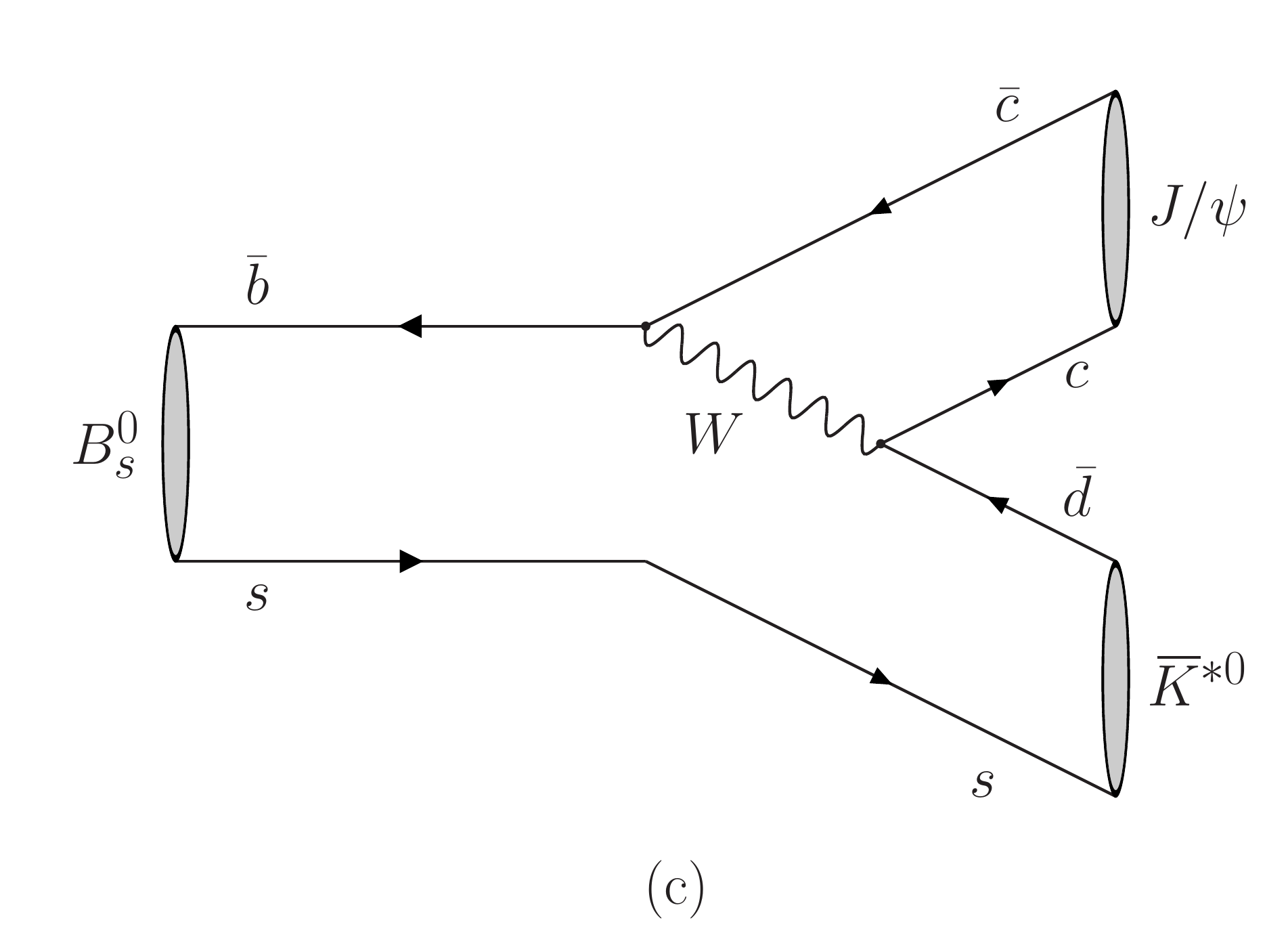}
\hfill
\includegraphics[width=0.49\textwidth]{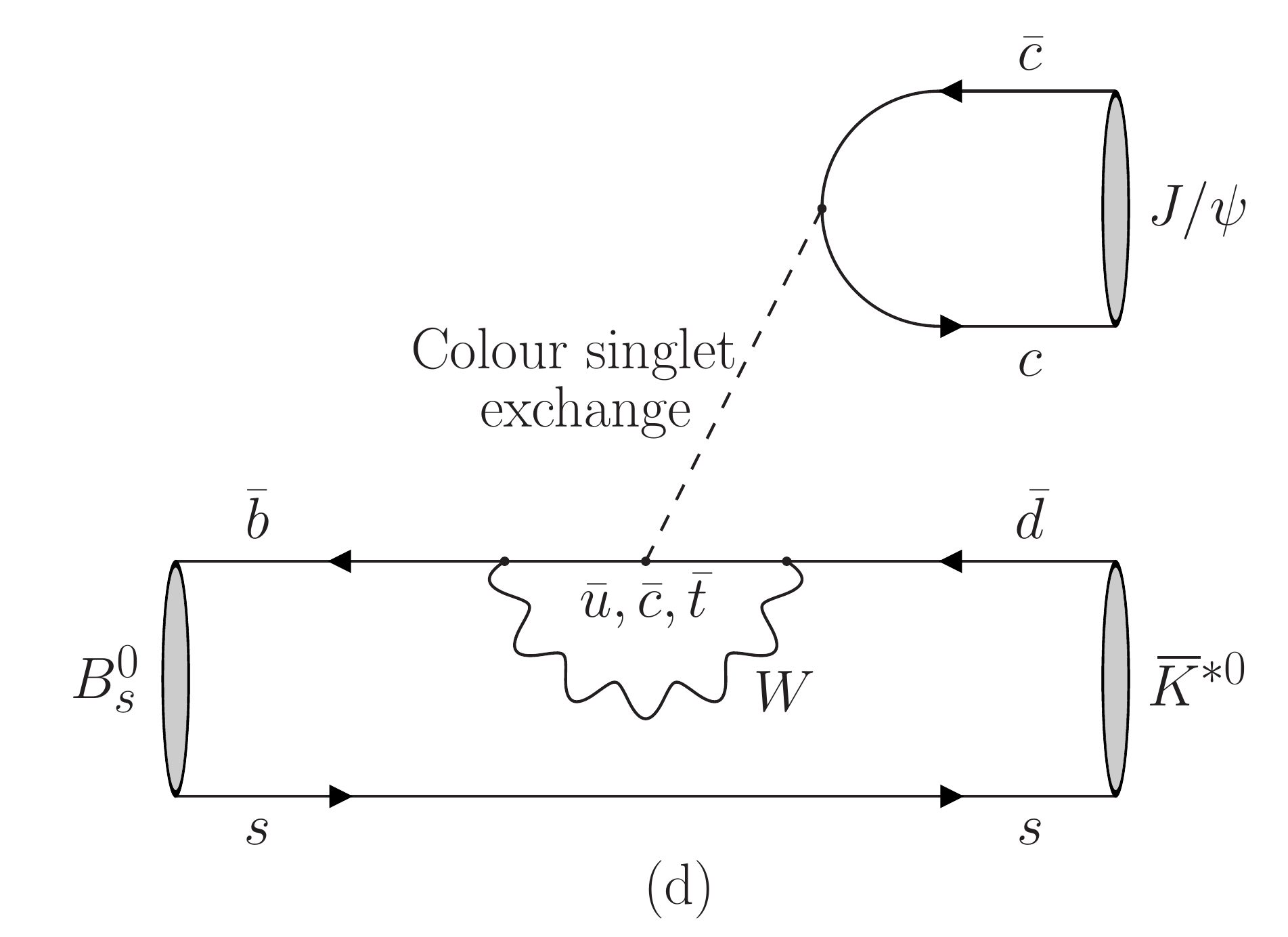}
\caption{Decay topologies contributing to the \BsJphi\ channel (a,b) and $\BsJKst$ channel (c,d).
The tree diagrams (a,c) are shown on the left and the penguin diagrams (b,d) on the right.}
\label{fig:diagBsJkst}
\end{figure}

The \CP-violating phase \phis arises in the interference between the  amplitudes of \Bs mesons decaying via $b\to c\bar{c}s$ transitions to \CP eigenstates directly and those decaying after  oscillation.
The phase \phis can be measured using the decay \BsJpsiPhi. 
Within the Standard Model (SM), and ignoring penguin contributions to the decay, \phis is predicted to be $-2\beta_s$, with \mbox{$\beta_s \equiv  \arg (-V_{cb}V_{cs}^*/V_{tb}V_{ts}^*)$,} where $V_{ij}$ are elements of the CKM matrix~\cite{Kobayashi:1973fv}. 
The phase \phis is a sensitive probe of dynamics beyond the SM (BSM) since it 
has a very small theoretical uncertainty and BSM processes can contribute to \Bs-\Bsb\ mixing~\cite{Altmannshofer:2009ne,Altmannshofer:2007cs,Buras:2009if,Chiang:2009ev}.
Global fits to experimental data, excluding the direct measurements of \phis, give \mbox{$-2\beta_s = -0.0363 \pm 0.0013\rad$~\cite{Charles:2015gya}.}
 The current world average value is $\phis = -0.015 \pm 0.035\rad$~\cite{HFAG}, dominated by the LHCb measurement reported in Ref.~\cite{LHCb-PAPER-2014-059}.
In the SM expectation of \phis~\cite{Charles:2015gya}, additional contributions to the leading $b\to c\bar{c}s$ tree Feynman diagram, as shown in \figref{fig:diagBsJkst}, are assumed to be negligible. However, the shift in \phis\ due to these contributions, called hereafter ``penguin pollution'', is difficult to compute due to the non-perturbative nature of the quantum chromodynamics (QCD) processes involved. This penguin pollution
must be measured or limited before using the \phis\ measurement in searches for BSM effects, since a shift in this phase caused by penguin diagrams is
possible. Various methods to address this problem have been
proposed~\cite{Fleischer:1999zi, Fleischer:2006rk, Faller:2008gt, Liu:2013nea, DeBruyn:2014oga, Frings:2015eva}, and LHCb has recently
published upper limits on the size of the penguin-induced phase shift using $\Bd\to \jpsi \rho^0$\ decays~\cite{LHCb-PAPER-2014-058}.

Tree and penguin diagrams contributing to both \BsJphi\ and \BsJpsiKst decays are shown in \figref{fig:diagBsJkst}.
In this paper, the penguin pollution in \phis is investigated using \BsJKst decays\footnote{Charge conjugation is implicit throughout this paper, unless otherwise specified.}, with $\jpsi\to\mu^+\mu^-$ and $\Kstarzb\to K^-\pi^+$, following the
method first proposed in Ref.~\cite{Fleischer:1999zi} for the $\Bd\to \jpsi \rho^0$ decay and later also discussed for the $\BsJKst$ decay in Refs.~\cite{Faller:2008gt, DeBruyn:2014oga}.
This approach requires the measurement of the branching
fraction, direct \CP asymmetries, and polarisation fractions of the \BsJpsiKst decay. The measurements use data from
proton-proton ($pp$) collisions recorded with the LHCb detector corresponding to 3.0\invfb of integrated luminosity,
of which 1.0 (2.0)\invfb was collected in 2011 (2012) at a centre-of-mass energy of 7 (8)\TeV. The LHCb collaboration previously reported
a measurement of the branching fraction and the polarisation fractions using data
corresponding to 0.37\invfb\ of integrated luminosity~\cite{LHCb-PAPER-2012-014}. 

The paper is organised as follows: a description of the LHCb detector, reconstruction and simulation software is given in \secref{sec:exp_setup}, the selection of the \BsJKst\ signal candidates and the \BdJKst\ control channel are presented in \secref{sec:selection} and the treatment of background in \secref{sec:peaking}.
The $\jpsi K^-\pi^+$ invariant mass fit is detailed in \secref{sec:sWeighting}. 
The angular analysis and \CP asymmetry measurements, both performed on weighted distributions where the background is statistically subtracted using the \splot technique~\cite{Pivk:2004ty}, are detailed in \secref{sec:angles}. 
The measurement of the branching fraction is explained in \secref{sec:normalisation}. 
The evaluation of systematic uncertainties is described in \secref{sec:syst} along with the results, and in \secref{sec:penguins} constraints on the penguin pollution are evaluated and discussed.

\section{Experimental setup}
\label{sec:exp_setup}

The \lhcb detector~\cite{LHCb-DP-2014-002,Alves:2008zz} is a single-arm forward
spectrometer covering the \mbox{pseudorapidity} range $2<\eta <5$,
designed for the study of particles containing \bquark or \cquark
quarks. The detector includes a high-precision tracking system
consisting of a silicon-strip vertex detector surrounding the $pp$
interaction region, a large-area silicon-strip detector located
upstream of a dipole magnet with a bending power of about
$4{\rm\,Tm}$, and three stations of silicon-strip detectors and straw
drift tubes placed downstream of the magnet.
The tracking system provides a measurement of momentum, \ptot, of charged particles with
a relative uncertainty that varies from 0.5\% at low momentum to 1.0\% at 200\gevc.
The minimum distance of a track to a primary vertex, the impact parameter, is measured with a resolution of (15+29$/\pt$)$\,\mum$,
where \pt is the component of the momentum transverse to the beam, in\,\gevc.
Different types of charged hadrons are distinguished using information
from two ring-imaging Cherenkov detectors. 
Photons, electrons and hadrons are identified by a calorimeter system consisting of
scintillating-pad and preshower detectors, an electromagnetic
calorimeter and a hadronic calorimeter. Muons are identified by a
system composed of alternating layers of iron and multiwire
proportional chambers.

The online event selection is performed by a trigger, 
which consists of a hardware stage, based on information from the calorimeter and muon
systems, followed by a software stage, which applies a full event
reconstruction.
In this analysis, candidates are first required to pass the hardware trigger,
  which selects muons with a transverse momentum $\pt>1.48\gevc$ 
  in the 7\tev data or $\pt>1.76\gevc$ in the 8\tev data.
  In the subsequent software trigger, at least
  one of the final-state particles is required to have both
  $\pt>0.8\gevc$ and impact parameter larger than 100$\,\mum$ with respect to all
  of the primary $pp$ interaction vertices~(PVs) in the
  event. Finally, the tracks of two or more of the final-state
  particles are required to form a vertex that is significantly
  displaced from any PV.
 Further selection requirements are applied offline  in order to increase the signal purity. 

  In the simulation, $pp$ collisions are generated using \pythia~\cite{Sjostrand:2006za,Sjostrand:2007gs}
  with a specific \lhcb configuration~\cite{LHCb-PROC-2010-056}.  Decays of hadronic particles
 are described by \evtgen~\cite{Lange:2001uf}, in which final-state
 radiation is generated using \photos~\cite{Golonka:2005pn}. The
 interaction of the generated particles with the detector, and its response,
 are implemented using the \geant toolkit~\cite{Allison:2006ve,  Agostinelli:2002hh} as described in
 Ref.~\cite{LHCb-PROC-2011-006}.

\section{Event selection}
\label{sec:selection}

The selection of \BsJpsiKst candidates consists of two steps: a preselection consisting of discrete cuts, followed by
a specific requirement on a boosted decision tree with gradient boosting (BDTG)~\cite{Breiman,AdaBoost} to suppress
combinatorial background. 
All charged particles are required to have a transverse momentum in excess of $0.5\gevcc$ and to be positively identified as
muons, kaons or pions. 
The tracks are fitted to a common vertex which is required to be of good quality and significantly displaced from any PV in the event. 
The flight direction can be described as a vector between the \Bs production and decay vertices; the cosine of the angle between this vector and the \Bs momentum vector is required to be greater than $0.999$. 
Reconstructed invariant masses of the \Jpsi and \Kstarzb candidates are required to be in the
ranges $2947 < m_{\mu^+\mu^-} < 3247 \mevcc$ and $826 < m_{K^-\pi^+} < 966 \mevcc$.
The \Bs\ invariant mass is reconstructed by constraining the \jpsi candidate to its nominal mass~\cite{PDG2014}, and is
required to be in the range $5150 < m_{\jpsi K^- \pi^+} < 5650 \mevcc$. 

The training of the BDTG is performed independently for 2011 and 2012 data, using information from the \Bs candidates: 
time of flight, transverse momentum, impact parameter with respect to the production vertex and $\chi^2$ of the
decay vertex fit.  
The data sample used to train the BDTG uses less stringent particle identification requirements. 
When training the BDTG, simulated \BsJpsiKst events are used to represent the signal, while candidates
reconstructed from data events with $\jpsi K^{-}\pi^{+}$ invariant mass above $5401 \mevcc$ are used to represent the background. The optimal threshold for the BDTG is chosen independently for 2011 and 2012 data and maximises the effective signal
yield. 
\section{Treatment of peaking backgrounds}
\label{sec:peaking}
After the suppression of most background with particle identification criteria, simulations show residual contributions from the backgrounds \mbox{\LbJpsipK}, \BsJpsiKK, \BsJpsipipi, and \BdJpsipipi. 
The invariant mass distributions of misidentified \BdJpsipipi and \BsJpsipipi
events peak near the \BsJpsiKpi signal peak due to the effect of a wrong-mass hypothesis, and the misidentified \BsJpsiKK candidates are located in the vicinity of the \BdJpsiKpi signal peak.
It is therefore not possible to separate such background from signal using information based solely on
the invariant mass of the $\Jpsi \Km\pip$ system. 
Moreover the shape of the reflected invariant mass distribution is sensitive to the daughter particles momenta. 
Due to these correlations it is difficult to add the \bJpsihh (where $h$ is either a pion, a kaon or a proton) misidentified backgrounds as extra modes to the fit to the invariant mass distribution. 
Instead, simulated events are added to the data sample with negative weights in order to cancel
the contribution from those peaking backgrounds, as done previously in Ref.~\cite{LHCb-PAPER-2014-059}.
Simulated \bJpsihh events are generated using a phase-space model, and then weighted 
on an event-by-event basis using the latest amplitude analyses of the decays \LbJpsipK~\cite{LHCb-PAPER-2015-029}, 
\BsJpsiKK~\cite{LHCb-PAPER-2012-040}, \BsJpsipipi~\cite{LHCb-PAPER-2013-069}, and \BdJpsipipi~\cite{LHCb-PAPER-2014-012}. 
The sum of weights of each decay mode is normalised such that the injected simulated events cancel out the expected yield in data of the specific background decay mode.

In addition to \LbJpsipK and \BJpsihh decays, background from \LbJpsippi is also expected. 
However, in Ref.~\cite{LHCb-PAPER-2014-020} a full amplitude analysis was not performed. For this reason, as well as the fact that the \Lb decays have broad mass distributions, the contribution is explicitly included in the mass fit described in the next section. Expected yields for both \BJpsihh and \LbJpsiph background decays are given in \tabref{tab:peakingSummary}. 
\begin{table}
\begin{center}
\caption{Expected yields of each background component in the signal mass range.}
\begin{tabular}{l c c}
\hline
Background sources & 2011 data & 2012 data \\
\hline
\BdJpsipipi & $\phantom{0}51 \pm 10\phantom{0}$ & $115 \pm 23\phantom{0}$ \\
\BsJpsipipi & $\phantom{0}9.3 \pm 2.1\phantom{0}$ & $25.0\pm 5.4\phantom{0}$\\
\BsJpsiKK & $10.1 \pm 2.3\phantom{0}$ & $19.2 \pm 4.0\phantom{0}$ \\
\LbJpsipK & $\phantom{0}36 \pm 17\phantom{0}$ & $\phantom{0}90 \pm 43\phantom{0}$ \\
\LbJpsippi & $13.8 \pm 5.3\phantom{0}$ & $27.3 \pm 9.0\phantom{0}$ \\
\hline
\end{tabular}
\label{tab:peakingSummary}
\end{center}
\end{table}

\section{Fit to the invariant mass distribution}
\label{sec:sWeighting}

After adding simulated \BdJpsipipi, \BsJpsipipi, \BsJpsiKK, and \mbox{\LbJpsipK} events with negative weights, the remaining sample consists of \mbox{\BdJpsiKpi}, \BsJpsiKpi, \LbJpsippi decays, and combinatorial background.
These four modes are statistically disentangled through a fit to the $\jpsi \Km\pip$ invariant mass. The combinatorial background is described by an exponential distribution, the \LbJpsippi decay by the Amoroso distribution~\cite{Amoroso} and the \Bd and \Bs signals by the double-sided Hypatia distribution~\cite{Santos:2013gra},
\begin{multline}
I(m,\mu,\sigma,\lambda,\zeta,\beta,a_1,a_2,n_1,n_2) \propto\\
\begin{cases}
\frac {A}{\left(B + m-\mu \right)^{n_1}}\phantom{,,,,,,,,,,,,,,,,,,,,,,,,,,,,,,,,,,,,,,,,-}\text{if } m - \mu < -a_1\sigma\,,\\
\frac {C}{\left(D + m-\mu \right)^{n_2}}\phantom{,,,,,,,,,,,,,,,,,,,,,,,,,,,,,,,,,,,,,,,,,,,,}\text{if } m - \mu > a_2\sigma\,,\\
\left((m-\mu)^{2} + \delta^{2}\right)^{\frac{1}{2} \lambda - \frac{1}{4}} e^{\beta (m-\mu)} K_{\lambda - \frac{1}{2}}\left(\alpha \sqrt{(m-\mu)^{2} + \delta^{2}}\right)\phantom{,,,,,,}\text{otherwise}\,,\phantom{,,}\\
\end{cases}
\end{multline}
where $K_{\nu}(z)$ is the modified Bessel function of the second kind, $\delta\equiv\sigma\sqrt{\frac{\zeta\,
K_\lambda(\zeta)}{K_{\lambda+1}(\zeta)}}$, \mbox{$\alpha\equiv\frac{1}{\sigma}\sqrt{\frac{\zeta\, K_{\lambda+1}(\zeta)}{K_\lambda(\zeta)}}$}, and
$A,B,C,D$ are obtained by imposing continuity and differentiability. This function is chosen because the
event-by-event uncertainty on the mass has a dependence on the particle momenta. 
The estimate of the number of \BdJpsiKpi decays lying under the \Bs peak is very sensitive to the modelling of the tails of the \Bd peak.
The fitted fraction is in good agreement with the estimate from simulation.

In the fit to data, the mean and resolution parameters of both the \Bs and \Bd Hypatia functions are free to vary. All the remaining parameters, namely $\lambda$, $a_1$, $n_1$, $a_2$ and $n_2$, are fixed to values determined from fits to \Bs and \Bd simulated events.
All the \LbJpsippi shape parameters are fixed to values obtained from fits to simulated \LbJpsippi events, while the exponent of the combinatorial background is free to vary. 

Due to the small expected yield of \LbJpsippi decays compared to those of the other modes determined in the fit to data, and to the broad distribution of \LbJpsippi decays across the $\jpsi \Km\pip$ invariant mass spectrum, its yield is included in the fit as a Gaussian constraint using the expected number of events and its uncertainties, as shown in \tabref{tab:peakingSummary}.

\begin{figure}[t]
\center
\begin{tabular}{c}
	\includegraphics[height= 7.8cm] {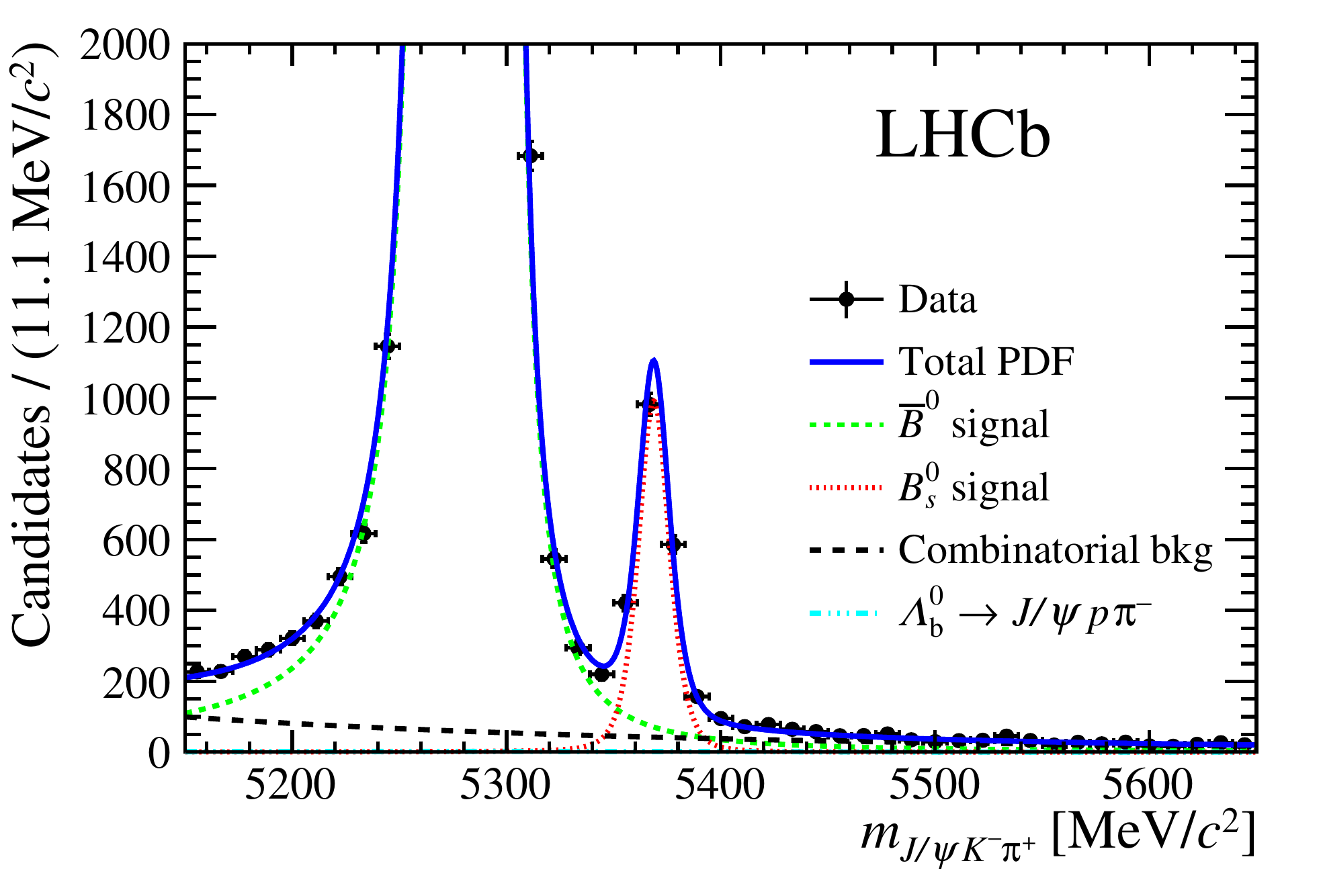} 
\end{tabular}
\caption{The $\jpsi \Km\pip$ invariant mass distribution with the sum of the fit projections in the 20 \mkpi and \cosTmu bins. 
Points with error bars show the data. 
The projection of the fit result is represented by the solid blue line, and the contributions from the different components are detailed in the legend. 
At this scale the contribution of the \LbJpsippi is barely visible. 
All the other peaking background components are subtracted as described in the text. \label{fig:massProjectionsAllBins}}
\end{figure}

From studies of simulated (MC) samples, it is found that the resolution of \Bs and \Bd mass peaks depends on both \mkpi and \cosTmu, where 
\thetamu is one of the helicity angles used in the angular analysis as defined in \secref{sec:angles}. 
The fit to the $\jpsi \Km\pip$ invariant mass spectrum, including the evaluation of the \sweights, is performed separately in twenty bins, corresponding to four \mkpi bins of $35\mevcc$ width, and five equal bins in \cosTmu. 
The overall \Bs and \Bd yields are obtained from the
sum of yields in the twenty bins, giving
\begin{align}\label{eqn:yieldBd}
N_{\Bd} &= 208656  \pm  462\:(\text{stat}) ^{+ 78	}_{- 76}\:(\text{syst})  \,,\\\label{eqn:yieldBs}
N_{\Bs} &= \phantom{00}1808  \pm   51\:(\text{stat}) ^{+ 38	}_{- 33}\:(\text{syst}) \,,
\end{align}
where the statistical uncertainties are obtained from the quadratic sum of the uncertainties determined in each of the individual fits. Systematic uncertainties are discussed in \secref{sec:syst}.
The correlation between the \Bd and \Bs yields in each bin are found to be smaller than $4\%$. 
The ratio of the \Bs and \Bd yields is found to be $N_{\Bs}/N_{\Bd} = (8.66  \pm  0.24\:(\text{stat})^{+ 0.18}_{- 0.16}\:(\text{syst})) \times 10^{-3}$. 
Figure~\ref{fig:massProjectionsAllBins} shows the sum of the fit results for each bin, overlaid on the $\jpsi \Km\pip$ mass spectrum for the selected data sample.

\section{Angular analysis\label{sec:angles}}

\subsection{Angular formalism}\label{sec:angformalism}
\begin{figure}[b]
  \centering
  \includegraphics[width=1.0\textwidth]{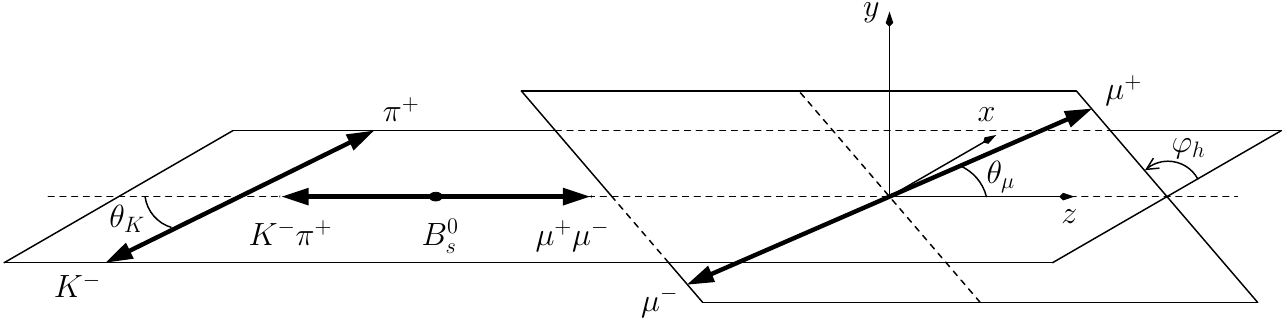}
  \caption{\small Representation of helicity angles as discussed in the text.}
  \label{fig:helicity}
\end{figure}

This analysis uses the decay angles defined in the helicity basis.
The helicity angles are denoted by $(\thetaK, \thetamu, \phihel)$, as shown in~\figref{fig:helicity}.
The polar angle \thetaK{} (\thetamu{}) is the angle between the kaon ($\mu^+$) momentum and the direction opposite to the \Bs{} momentum in the $\Km\pip$ ($\mu^+\mu^-$) centre-of-mass system.
The azimuthal angle between the $\Km\pip$ and $\mu^+\mu^-$ decay planes is \phihel.
The definitions are the same for \Bs\ or \Bsb\ decays. They are also the same for \BdJKst\ decays.

The shape of the angular distribution of \BsJpsiKst decays is given by Ref.~\cite{SheldonAndLiming},
\begin{equation}
\label{eq:angPdf}
\frac{\text{d}\Gamma(\theta_K,\thetamu,\phihel)}{\text{d}\Omega} \propto \sum_{\alpha_{\mu} = \pm1} \Bigg|
\sum_{\lambda,J}^{|\lambda|<J}\sqrt{\frac{2J+1}{4\pi}}\mathcal{H}_{\lambda}^{J}
e^{-\text{i}\lambda\phihel}d_{\lambda,\alpha_{\mu}}^{1}(\thetamu)d_{-\lambda,0}^{1}(\thetaK) \Bigg|^2\:,
\end{equation}
\noindent where $\lambda=0,\pm1$ is the \Jpsi helicity, $\alpha_{\mu}=\pm1$ is the helicity
difference between the muons, $J$ is the spin of the $\Km\pip$ system, $\mathcal{H}$ are the helicity
amplitudes, and $d$ are the small Wigner matrices.

The helicity amplitudes are rotated into transversity amplitudes, which correspond to final 
$P$~eigenstates,
\begin{align}
A_{\rm S} &= \mathcal{H}_0^0    \,, \label{eq:helAmp1}\\
A_{0} &= \mathcal{H}_0^1   \,, \label{eq:helAmp2}\\
A_{\parallel} &=  \frac{1}{\sqrt{2}}({\mathcal H}_+^1 + {\mathcal H}_-^1)  \,, \label{eq:helAmp3}\\
A_{\perp} &= \frac{1}{\sqrt{2}}({\mathcal H}_+^1 - {\mathcal H}_-^1)   \,. \label{eq:helAmp4}
\end{align}
The distribution in \equref{eq:angPdf} can be written as the sum of ten angular terms, four corresponding to the square of the transversity amplitude of each final state polarisation, and six corresponding to the cross terms describing interference among the final polarisations.

The modulus of a given transversity amplitude, $A_x$, is written as $|A_x|$, and its phase as $\delta_x$. 
The convention $\delta_0 = 0$ is used in this paper.
The \pwave polarisation fractions are $f_i = |A_i|^2 / ( |A_0|^2 + |A_\parallel|^2 + |A_\perp|^2 )$, with $i=0, \parallel, \perp$ and 
the \swave fraction is defined as $F_{\rm S} = |A_{\rm S}|^2 / ( |A_0|^2 + |A_\parallel|^2 + |A_\perp|^2 + |A_{\rm S}|^2)$. 
The distribution of the \CP-conjugate decay is obtained by flipping the sign of the interference terms
which contain $|A_{\perp}|$. 
For the \CP-conjugate case, the amplitudes are denoted as $\overline{A}_i$. Each $A_i$ and the corresponding $\overline{A}_i$ are related
through the \CP asymmetries, as described in \secref{sec:cpv}.

\subsection{Partial-wave interference factors}
\label{sec:CSP}

\begin{figure}[h!]
\center
	\includegraphics[height= 5.4cm] {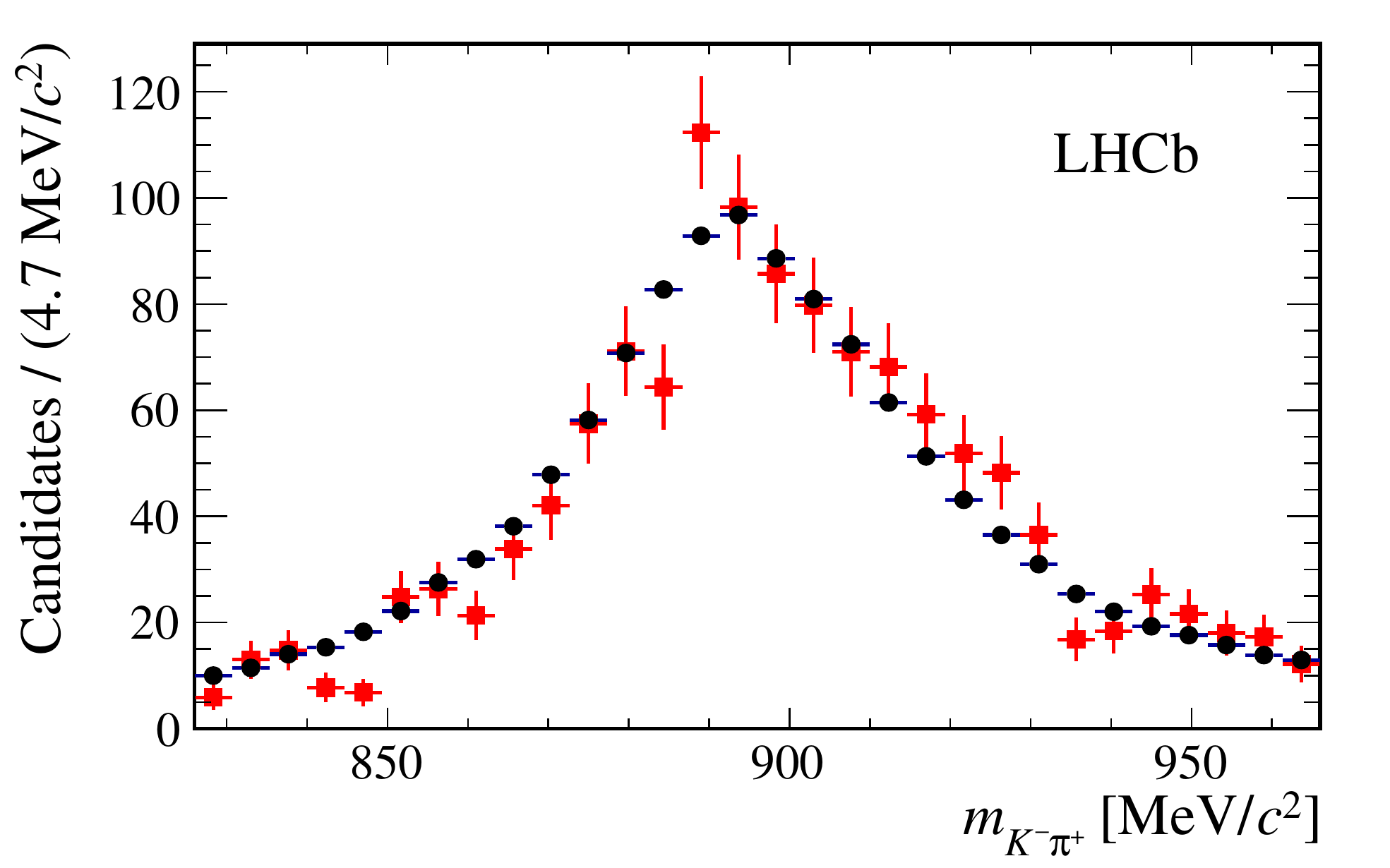}\\
\caption[$K\pi$ invariant mass distributions. R]{Efficiency corrected \mkpi distribution for \Bs shown in squares (red) and \Bdb shown in circles (black) using \sweights computed from the maximum likelihood fit to the $\jpsi \Km\pip$ invariant mass spectrum.
\label{fig:Kpi_invMass_BsBd_w_and_wo_acceptance}}
\end{figure}

In the general case, the transversity amplitudes of the angular model depend on the $\Km\pip$ mass (\mkpi). This variable is limited to be inside a window of $\pm$70\MeVcc around the \Kstarzb mass. 
Figure~\ref{fig:Kpi_invMass_BsBd_w_and_wo_acceptance} shows the efficiency-corrected \mkpi spectra for \Bs and \Bd using the nominal sets of \sweights.

In order to account for the \mkpi dependence while keeping the framework of an angular-only
analysis, a fit is performed simultaneously in the same four \mkpi bins defined in \secref{sec:sWeighting}.
Different values of the parameters $|A_{\rm S}|^{2}$ and $\delta_{\rm S}$ are allowed for
each bin, but the angular distribution still contains mass-dependent terms associated with the interference between partial-waves.
If only the \swave and \pwave are considered, such interference terms correspond
to the following complex integrals,
\begin{equation}
\label{eq:CSP}
\frac{\int_{m_{K\pi}^L}^{m_{K\pi}^H} { \mathcal{P} \times \mathcal{S}^*\:{\Phi}\:\varepsilon_{m}(m_{K\pi})}  \:\:\text{d}
m_{K\pi}}
{\sqrt{\int_{m_{K\pi}^L}^{m_{K\pi}^H} { |\mathcal{P}|^2\:{\Phi}\:\varepsilon_{m}(m_{K\pi})}  \:\:\text{d}
m_{K\pi} \int_{m_{K\pi}^L}^{m_{K\pi}^H} { |\mathcal{S}|^2\:{\Phi}\:\varepsilon_{m}(m_{K\pi})}  \:\:\text{d}
m_{K\pi}}}
 = C_{\rm SP} e^{-\text{i} \theta_{\rm SP}} \,,
\end{equation}
where $m_{K\pi}^{L(H)}$ is the lower (higher) limit of the bin, $\varepsilon_{m}(m_{K\pi})$ is the acceptance for a $\Km\pip$ candidate with mass $m_{K\pi}$ (see \appref{sec:angularAcc} for a discussion on the angular acceptance), $\Phi$ stands for the phase space, and $\mathcal{P}$ ($\mathcal{S}$) is the \pwave (\swave) propagator. The phase space term is computed as
\begin{equation}
\Phi=\frac{p\:q}{m_{K\pi}^{2}},
\end{equation}
where $p$ denotes the $\Kstarzb$ momentum in the $B_{s}^{0}$ rest frame and $q$ refers to the $K^{-}$ momentum in the $\Kstarzb$ rest frame.

The phase $\theta_{\rm SP}$ is included in the definition of $\delta_{\rm S}$ but the $C_{\rm SP}$ factors, corresponding to real numbers in the interval $[0,1]$, have to be computed and input to the angular fit. 
The contribution of \dwave ($J=2$) in the \mkpi range considered is expected to be negligible. 
Therefore the nominal model only includes \swave and \pwave. To determine the systematic uncertainty due to possible \dwave contributions, $C_{\rm SD}$ and $C_{\rm PD}$ factors are also computed, using analogous expressions to that given in \equref{eq:CSP}. The $C_{ij}$ factors are calculated by evaluating numerically the integrals using the propagators outlined below, and are included as fixed parameters in the fit. 
A systematic uncertainty associated to the different possible choices of the propagator models is afterwards evaluated.

The \swave propagator is constructed using the LASS parametrisation~\cite{Aston1988493},
consisting of a linear combination of the $K^{*}_{0}(1430)^{0}$ resonance with a non-resonant term,
coming from elastic scattering. The \pwave is described by a combination of the
$K^{*}(892)^{0}$ and $K^{*}_{1}(1410)^{0}$ resonances using the isobar model~\cite{Herndon:1973yn}, and the \dwave is assumed to come from the
$K^{*}_{2}(1430)^{0}$ contribution. Relativistic Breit-Wigner functions, multiplied by angular momentum barrier factors, are used to parametrise the
different resonances. \tabref{tab:CSPfactors} contains the computed $C_{\rm SP}$, $C_{\rm SD}$ and $C_{\rm PD}$
factors. 

\begin{table}[t]
  \begin{center}
  \caption{The $C_{\rm SP}$, $C_{\rm SD}$ and $C_{\rm PD}$ factors calculated in each of the four \mkpi bins around the \Kstarzb peak.}
    \begin{tabular}{c c c c c}
      \hline
      Bin & \mkpi range (\MeVcc) & $C_{\rm SP}$ & $C_{\rm SD}$ & $C_{\rm PD}$ \\
      \hline
      0 & [826,\,861] & 0.968 $\pm$ 0.017 & 0.9968 $\pm$ 0.0030 & 0.9827 $\pm$ 0.0048 \\
      1 & [861,\,896] & 0.931 $\pm$ 0.012 & 0.9978 $\pm$ 0.0021 & 0.9402 $\pm$ 0.0048 \\
      2 & [896,\,931] & 0.952 $\pm$ 0.012 & 0.9983 $\pm$ 0.0016 & 0.9421 $\pm$ 0.0056 \\
      3 & [931,\,966] & 0.988 $\pm$ 0.011 & 0.9986 $\pm$ 0.0012 & 0.9802 $\pm$ 0.0066 \\
      \hline
    \end{tabular}
  \label{tab:CSPfactors}
  \end{center}
\end{table}

\subsection{\CP asymmetries}
\label{sec:cpv}

The direct \CP violation asymmetry in the $B^0_{(s)}$ decay rate to the final state $f_{(s)\,i}$, with $f_{s,i} = \jpsi (K^-\pi^+)_i$ and $f_i=\jpsi (K^+\pi^-)_i$, is defined as 
\begin{equation}
\label{decayrates}
A^{\CP}_i (B^0_{(s)} \to f_{(s)\,i}) = \frac{|\overline{A}_{(s)\,i}|^2 -|A_{(s)\,i}|^2}{| \overline{A}_{(s)\,i}|^2 + |A_{(s)\,i}|^2} \,,
\end{equation}
where $A_{(s)\,i}$ are the transversity amplitudes defined in \secref{sec:angformalism} and the additional index $s$ is used to distinguish the \Bs and the \Bd-meson.  
The index $i$ refers to the polarisation of the final state ($i = 0,\parallel,\perp, {\rm S}$) and is dropped in the rest of this section, for clarity. 

The raw \CP asymmetry is expressed in terms of the number of observed candidates by
\begin{equation}
A^{\CP}_{\rm raw}(B^0_{(s)} \to f_{(s)}) = \frac{N^{\rm obs}(\overline{f}_{(s)}) -N^{\rm obs}(f_{(s)}) }{
N^{\rm obs}(\overline{f}_{(s)}) + N^{\rm obs}(f_{(s)})} \,.
\label{acp_mes}
\end{equation}
Both asymmetries in Eq.~\ref{decayrates} and Eq.~\ref{acp_mes} are related by~\cite{LHCB-PAPER-2011-029}
\begin{equation}
A^{\CP}(B^0_{(s)} \to f_{(s)}) \simeq A^{\CP}_{\rm raw}(B^0_{(s)} \to f_{(s)}) - \zeta_{(s)}
A_{\rm D}(f) - \kappa_{(s)}A_{\rm P}(B^0_{(s)})  \,,
\end{equation}
where $ A_{\rm D}(f)$ is the detection asymmetry, defined as in Eq.~\eqref{eq:detAsymm}, $A_{\rm P}(B^0_{(s)})$ is the $B^0_{(s)}$$-$$\Bbar^0_{(s)}$ production asymmetry, defined as in Eq.~\eqref{eq:prodAsymm}, 
$\zeta_{(s)}= +1(-1)$ and $\kappa_{(s)}$ accounts for the dilution due to
$B^0_{(s)}$$-$$\Bbar^0_{(s)}$ oscillations~\cite{LHCb-PAPER-2013-018}. The $\kappa_{(s)}$ factor
is evaluated by
\begin{equation}
\kappa_{(s)} \!=\! \frac{\int_0^\infty \! e^{-\Gamma_{(s)} t} \! \cos \!\left( \Delta m_{(s)} t
\right )\! \varepsilon(t)\mathrm{d}t}{\int_0^\infty \!
e^{-\Gamma_{(s)} t} \! \cosh \!\left( \frac{\Delta \Gamma_{(s)}}{2} t \right )\!
\varepsilon(t)\mathrm{d}t}  \,,
\label{kappa}
\end{equation}
where  $\varepsilon(t)$ is the time-dependent acceptance function, assumed to be identical for the \BsJKst\ and \BdJKst\ decays.  
The symbols $\Gamma_{(s)}$ and $\Delta m_{(s)}$ denote the decay width and mass differences between the $B^0_{(s)}$ mass eigenstates. 

The $B^0_{(s)}$$-$$\Bbar^0_{(s)}$ production asymmetry is defined as
\begin{equation}\label{eq:prodAsymm}
A_{\text{P}}\left(B_{(s)}^0\right) \equiv \frac{\sigma\left(\Bbar_{(s)}^0\right) - \sigma\left(B_{(s)}^0\right)}{ \sigma \left(\Bbar_{(s)}^0\right) + \sigma\left(B_{(s)}^0\right) }\:,
\end{equation}
where $\sigma$ is the $B_{(s)}^0$ production cross-section within the LHCb acceptance. The production asymmetries reported in Ref.~\cite{LHCb-PAPER-2014-042} are reweighted in bins of $B_{(s)}^0$ transverse momentum to obtain
\begin{align}
A_{\text{P}}(\Bd) = & \left(-1.04 \pm 0.48\:(\text{stat}) \pm 0.14\:(\text{syst}) \right)\%\:, \nonumber \\ 
A_{\text{P}}(\Bs) = & \left(-1.64 \pm 2.28\:(\text{stat}) \pm 0.55\:(\text{syst}) \right)\%  \,.  \nonumber
\end{align}
The $\kappa_{(s)}$ factor in Eq.~\ref{kappa} is determined by fixing  $\Delta\Gamma_{(s)}$, $\Delta m_{(s)}$ and  $\Gamma_{(s)}$ to their world average values~\cite{PDG2014} and by fitting the decay time acceptance $\varepsilon(t)$ to the nominal data sample after applying the \Bd \sweights, in a similar way to Ref.~\cite{LHCb-PAPER-2014-053}. 
It is equal to 0.06$\%$ for \Bs decays, and 41$\%$ for \Bd.
This reduces the effect of the production asymmetries to the level of $10^{-5}$ for \BsJpsiKst and $10^{-3}$ for \BdJpsiKst decays. 

Other sources of asymmetries arise from the
different final-state particle interactions with the detector, event reconstruction and detector acceptance.
The detection asymmetry, $A_{\text{D}}(f)$, is defined in terms of the detection efficiencies of the final states, $\varepsilon^{\rm det}$, as 
\begin{equation}\label{eq:detAsymm}
A_{\text{D}}(f) \equiv \frac{\varepsilon^{\rm det}(\overline{f}) - \varepsilon^{\rm det}(f)}{\varepsilon^{\rm det}(\overline{f}) + \varepsilon^{\rm det}(f)} \,.
\end{equation}
%
The detection asymmetry, measured in bins of the \Kp momentum in Ref.~\cite{LHCB-PAPER-2014-013}, is weighted with the momentum distribution of the kaon from the \BJpsiKst decays to give
%
\begin{eqnarray}
A_{\text{D}}(\Bd) & = & \left(\phantom{-}1.12  \pm 0.55 \:(\text{stat}) \right)\% \,,\nonumber \\ 
A_{\text{D}}(\Bs) & = & \left({-}1.09 \pm 0.53 \:(\text{stat}) \right)\% \,. \nonumber
\end{eqnarray}

\section{Measurement of \BRof\BsJpsiKst}\label{sec:normalisation}

The branching fraction \BRof\BsJpsiKst is obtained by normalising to two different channels, \BsJpsiPhi and \BdJpsiKst, and then averaging the results.
The expression
\begin{multline}
\frac{\BRof\BsJpsiKst\times\BRof{\Kst\to K^+\pi^-}}{\BRof\BJpsiX\times\BRof{X \to h^+h^-}} = 
\frac{N_{\BsJpsiKst}}{N_{\BJpsiX}}\times\frac{\varepsilon_{\BJpsiX}}{\varepsilon_{\BsJpsiKst}}\times\frac{f_q}{f_s}\,,
\end{multline}
is used for the normalisation to a given \BJpsiX decay, 
where $N$ refers to the yield of the given decay, $\varepsilon$ corresponds to the 
total (reconstruction, trigger and selection) efficiency, and  $f_q =f_s(f_d)$ are the $B_s^0$(\Bd)-meson hadronisation fractions. 

The event selection of \BsJpsiPhi candidates consists of the same requirements as those for \BsJpsiKst candidates (see \secref{sec:selection}), with the exception that $\phi$ candidates are reconstructed in the $\Kp\Km$ state so there are no pions among the final state particles.
In addition to the other requirements, reconstructed $\phi$ candidates are required to have mass in the range $1000 < m_{K^-K^+} < 1040 \mevcc$ and to have a transverse momentum in excess of $1\gevcc$.

\subsection{Efficiencies obtained in simulation} 
A first estimate of the efficiency ratios is taken from simulated events, where the particle identification variables are calibrated using \Dstpm decays.
The efficiency ratios estimated from simulation, for 2011 (2012) data, are
$\varepsilon_{\BdJpsiKst}^{\rm MC}/\varepsilon_{\BsJpsiKst}^{\rm MC} = 0.929 \pm 0.012~(0.927 \pm 0.012)$
and \mbox{$\varepsilon_{\BsJpsiPhi}^{\rm MC}/\varepsilon_{\BsJpsiKst}^{\rm MC} = 1.991 \pm 0.025~(1.986 \pm 0.027)$}.

\subsection{Correction factors for yields and efficiencies}

The signal and normalisation channel yields obtained from a mass fit are affected by the presence
of a non-resonant \swave background as well as interference between \swave and \pwave components. Such interference would integrate to
zero for a flat angular acceptance, but not for experimental data that are subject to an angle-dependent
acceptance. In addition, the efficiencies determined in simulation correspond to events generated with an angular distribution different from that in data; therefore the angular integrated efficiency also needs to be modified with respect to simulation
estimates.
These effects are taken into account using a correction factor~$\omega$, which is the product of the
correction factor to the angular-integrated efficiency and the correction factor to the \pwave yield:
\begin{equation}
\frac{N_{\BsJpsiKst}}{N_{\BJpsiX}}\times\frac{\varepsilon_{\BJpsiX}}{\varepsilon_{\BsJpsiKst}} = 
 \frac{N_{\BsJpsiKst}}{N_{\BJpsiX}}\times\frac{\varepsilon_{\BJpsiX}^{\rm MC}}{\varepsilon_{\BsJpsiKst}^{\rm MC}}\times\frac{\omega_{\BJpsiX}}{\omega_{\BsJpsiKst}} \,,
\end{equation}
\noindent where $N_{\BsJpsiKst}$, $N_{\BJpsiX}$ are the yields obtained from the mass fits,
$\varepsilon_{\BJpsiX}^{\rm MC}, \varepsilon_{\BsJpsiKst}^{\rm MC}$ are the efficiencies obtained in
simulation, and $\omega$ is calculated as
\begin{equation}
\omega_{\BJpsiX} = \frac{F_{\BJpsiX}^X}{c_{\BJpsiX}} \,,
\end{equation}
\noindent where $F_{\BJpsiX}^X$ is the fraction of the \pwave $X$ resonance in a given $\BJpsiX$ decay (related to the presence of \swave and its interference with the \pwave), and $c_{\BJpsiX}$ is a correction to $\varepsilon_{\BJpsiX}^{\rm MC}$ due to the fact that the simulated values of the decay parameters differ slightly from those measured.
The values obtained for the $\omega$ correction factors are
\begin{align}
\omega_{\BsJpsiKst} &= 1.149 \pm 0.044 \,{\rm (stat)} \pm 0.018 \,{\rm (syst)}  \,, \nonumber \\
\omega_{\BdJpsiKst} &= 1.107 \pm 0.003 \,{\rm (stat)} \pm 0.038 \,{\rm (syst)}  \,, \nonumber \\
\omega_{\BsJpsiPhi} &= 1.013 \pm 0.002 \,{\rm (stat)} \pm 0.007 \,{\rm (syst)} \,. \nonumber
\end{align}

\subsection{Normalisation to \BsJpsiPhi}\label{sec:BF_norm_JpsiKK}

The study of penguin pollution requires the calculation of ratios of absolute amplitudes between \BsJpsiKst and \BsJpsiPhi. Thus, normalising \BRof\BsJpsiKst to \mbox{\BRof\BsJpsiPhi} is very useful. This normalisation is given by
\begin{equation}
\label{eq:norm_jpsiPhi}
\small
\frac{\BRof\BsJpsiKst}{\BRof\BsJpsiPhi} = \frac{N_{\BsJpsiKpi}}{N_{\BsJpsiKK}} \times \frac{\varepsilon_{\BsJpsiPhi}^{\rm MC}}{\varepsilon_{\BsJpsiKst}^{\rm MC}} \times \frac{\omega_{\BsJpsiPhi}}{\omega_{\BsJpsiKst}} \times \frac{\BRof{\phi\to K^+K^-}}{\BRof{\Kstarzb\to K^-\pi^+}} \,,
\end{equation}
where $\BRof{\Kstarzb\to K^-\pi^+}= 2/3$ and $\BRof{\phi\to K^+K^-} = (49.5 \pm 0.5)\%$~\cite{PDG2014}. Using $N_{\BsJpsiKpi}$ as given in \equref{eqn:yieldBs}, and $N_{\BsJpsiKK} = 58\,091~\pm~243~({\rm stat}) \pm 319~({\rm syst})$ as obtained from a fit to the invariant mass of selected \BsJpsiPhi candidates, where the signal is described by a double-sided Hypatia distribution and the combinatorial background is described by an exponential distribution, a value of
\begin{equation}
\label{eq:norm_jpsiPhi_result}
\frac{\BRof\BsJpsiKst}{\BRof\BsJpsiPhi} = \big(4.05 \pm 0.19 ({\rm stat}) \pm 0.13 ({\rm syst})\big)\% \nonumber \\
\end{equation}
is obtained.

\subsection{Normalisation to \BdJpsiKst}

The normalisation to \BdJpsiKst is given by
\begin{equation}
\label{eq:norm_jpsiKst}
\frac{\BRof\BsJpsiKst}{\BRof\BdJpsiKst} = \frac{N_{\BsJpsiKpi}}{N_{\BdJpsiKpi}}\times \frac{f_d}{f_s} \times \frac{\varepsilon_{\BdJpsiKst}^{\rm MC}}{\varepsilon_{\BsJpsiKst}^{\rm MC}} \times \frac{\omega_{\BdJpsiKst}}{\omega_{\BsJpsiKst}} \,,
\end{equation}
\noindent where $N_{\BdJpsiKpi}$ and $N_{\BsJpsiKpi}$ are given in \equref{eqn:yieldBd} and \equref{eqn:yieldBs}, respectively, and
\begin{equation}
\frac{\omega_{\BdJpsiKst}}{\omega_{\BsJpsiKst}} = 0.963 \pm 0.036\,  ({\rm stat}) \pm 0.031 \,({\rm syst}) \,, \nonumber \\
\end{equation}
\noindent resulting in a value of
\begin{equation}
\label{eq:norm_jpsiKst_result}
\frac{\BRof\BsJpsiKst}{\BRof\BdJpsiKst} = \left(2.99\pm 0.14 \, ({\rm stat}) \pm 0.12 \, ({\rm syst}) \pm 0.17\, (f_d/f_s)\right)\%  \,,  \\
\end{equation}
where the third uncertainty comes from the hadronisation fraction ratio \mbox{$f_d/f_s = 3.86 \pm 0.22$}~\cite{HFAG}.

\subsection{Computation of \BRof\BsJpsiKst}\label{sec:averageBsJpsiKst}

By multiplying the fraction given in \equref{eq:norm_jpsiKst_result} by the branching fraction of the decay \BdJpsiKst measured at Belle\footnote{The result from Belle was
chosen rather than the PDG average, since it is the only \BRof\BdJpsiKst measurement that subtracts \swave contributions.},
$(1.29 \pm0.05 \,({\rm stat}) \pm 0.13 \,({\rm syst)}) \times 10^{-3}$~\cite{Abe:2002haa}, and taking into account the difference in production rates for the \Bu{}\Bub and \Bd{}\Bdb pairs at the $\Upsilon(4S)$ resonance, i.e.\ $\Gamma(\Bu\Bub)/\Gamma(\Bd\Bdb) = 1.058 \pm 0.024$ \cite{HFAG}, the value
\begin{eqnarray}
\BR{\BsJpsiKst}_{d} & =  (3.95 \hspace{-0.3cm} & \pm 0.18 \, ({\rm stat}) \pm 0.16 \, ({\rm syst}) \pm 0.23 \, (f_d/f_s) \nonumber \\ 
                   &         & \pm 0.43 \,(\BRof\BdJpsiKst) )\times10^{-5} \, \nonumber
\end{eqnarray}
is obtained, where the fourth uncertainty arises from \mbox{$\BRof\BdJpsiKst$}.
A second estimate of this quantity is found via the normalisation to \BRof\BsJpsiPhi~\cite{LHCb-PAPER-2012-040}, 
 updated with the value of $f_d/f_s$ from Ref.~\cite{HFAG} to give $\BRof\BsJpsiPhi=(1.038\pm0.013\:(\text{stat})\pm 0.063\:(\text{syst})\pm 0.060\:(f_d/f_s))\times10^{-3}$, resulting in a value of
\begin{equation}
\BR{\BsJpsiKst}_{\phi} =  \left(4.20 \pm 0.20  \, \text{(stat)} \pm  0.13  \, \text{(syst)}  \pm 0.36 \, (\BRof\BsJpsiPhi)\right)\times 10^{-5} \,, \nonumber
\end{equation}
where the third uncertainty comes from $\BRof\BsJpsiPhi$. Both values
are compatible within uncorrelated systematic uncertainties and are combined, taking account of correlations, to give
\begin{equation}\label{eq:averagedBRvalue}
\BR{\BsJpsiKst} = \left(4.14 \pm 0.18 \, \text{(stat)} \pm  0.26 \, \text{(syst)} \pm 0.24 \, (f_d/f_s)\right)\times 10^{-5} \,,\nonumber
\end{equation}
which is in good agreement with the previous LHCb measurement~\cite{LHCb-PAPER-2012-014}, of $(4.4_{-0.4}^{+0.5} \pm 0.8) \times 10^{-5}$.

\section{Results and systematic uncertainties}
\label{sec:syst}

Section~\ref{sec:systAng} presents the results of the angular fit as well as the procedure used to estimate the systematic uncertainties, while in \secref{sec:systBR} the results of the branching fraction measurements and the corresponding estimated systematic uncertainties are discussed.

\subsection{Angular parameters and \CP asymmetries}\label{sec:systAng}

\begin{figure}[b!]
  \includegraphics[scale=0.8]{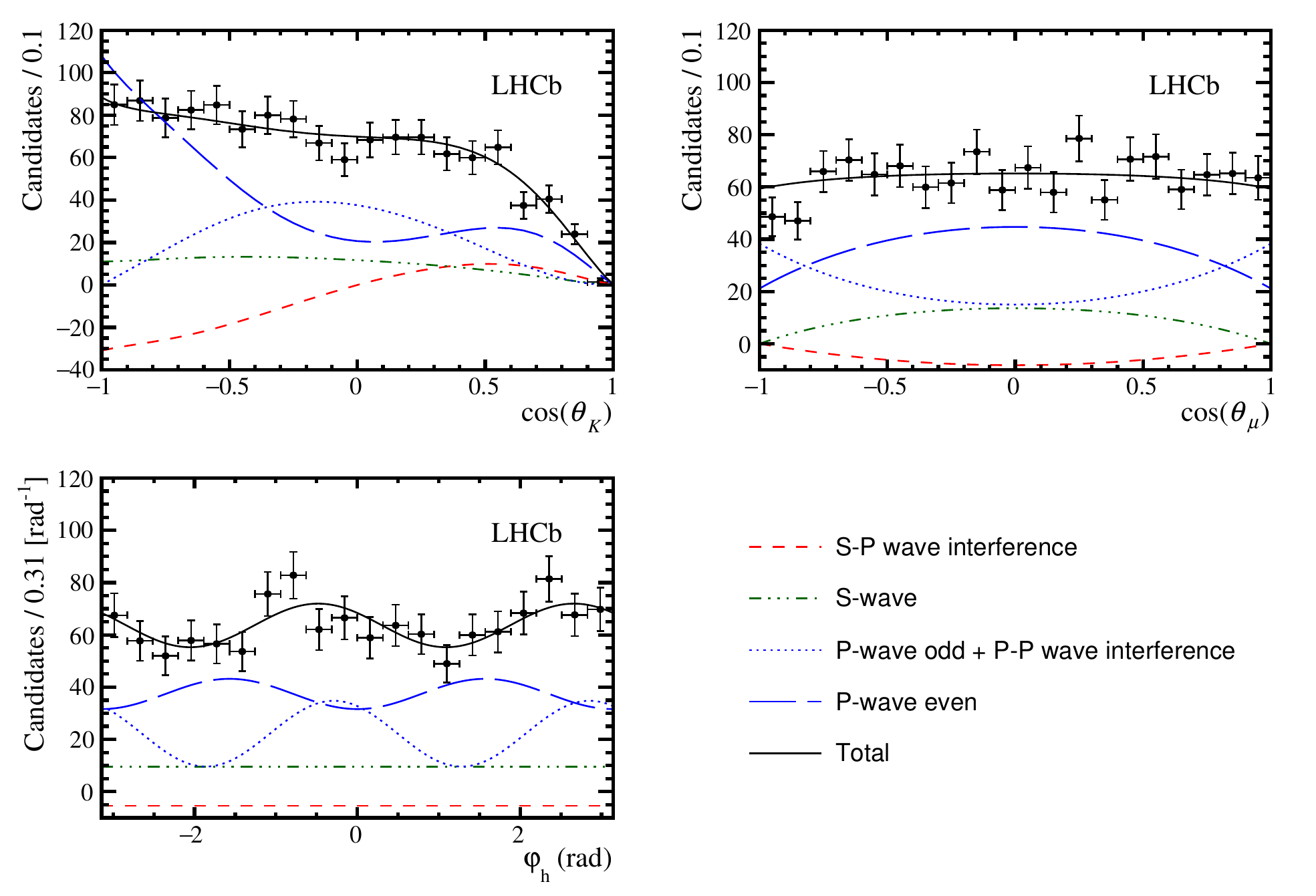}
  \caption{\small Fitted signal distributions compared with the weighted angular distributions with \Bs \sweights. Points with error bars show the data.
The projection of the fit result is represented by the solid black line, and the contributions from the different amplitude components are described in the legend.}
  \label{fig:angular_fit}
\end{figure}

The results obtained from the angular fit to the \BsJpsiKst events are given in \tabref{tab:systPwave} and \tabref{tab:systSwave} for the \mbox{\pwave} and \mbox{\swave} parameters, respectively. For comparison, the previous LHCb measurements~\cite{LHCb-PAPER-2012-014} of $f_0$ and $f_\|$ were $0.50 \pm 0.08 \pm 0.02$ and $0.19^{+0.10}_{-0.08} \pm 0.02$, respectively.
The angular distribution of the signal and the projection of the fitted distribution are shown in~\figref{fig:angular_fit}.
The statistical-only correlation matrix as obtained from the fit to data is given in \appref{App:CorrMatrix}.
The polarisation-dependent \CP asymmetries are compatible with zero, as expected in the SM. 
The polarisation fractions are in good agreement with the previous measurements~\cite{LHCb-PAPER-2012-014} performed on the same decay mode by the LHCb collaboration using data corresponding to an integrated luminosity of \mbox{0.37\invfb}.

Various sources of systematic uncertainties on the parameters of the angular fit are studied, as summarised in \tabref{tab:systPwave} and \tabref{tab:systSwave} for the \pwave and \swave parameters. 
Two classes of systematic uncertainties are defined, one from the angular fit model and another from the mass fit model. 
Since the angular fit is performed on the data weighted using the signal \sweights calculated from the fit to the $\jpsi K^-\pi^+$ invariant mass, biases on the mass fit results may be propagated to the \sweights and thus to the angular parameters. 
Overall, two sources of systematic uncertainties dominate: the angular acceptance and the  correlation between the $\jpsi K^-\pi^+$ invariant mass and \thetamu. 
\begin{table}[t!]
\begin{center}
\caption[]{Summary of the measured \BsJpsiKst~\pwave properties and their statistical and
systematic uncertainties. 
When no value is given, it means an uncertainty below $5\times10^{-4}$, except for the two phases, $\delpar$ (rad) and $\delperp$ (rad), in which case the uncertainty is below $5\times10^{-3}$.}
\small{
\begin{tabular}{lcccccccc}
\hline
\noalign{\vskip 0.02 in}

  &  $\fL$  & $\fpa$ & $\delpar$ & $\delperp$ & \ACPL & \ACPpa & \ACPpe  \\ [+0.04in]
\hline
\multirow{2}{*}{Fitted value} & \multirow{2}{*}{$0.497$} & \multirow{2}{*}{$0.179$} & \multirow{2}{*}{$-2.70$}& \multirow{2}{*}{$0.01$}& \multirow{2}{*}{$-0.048$}& \multirow{2}{*}{$0.171$} & \multirow{2}{*}{$-0.049$} \\ 
           & & & & & & &  \\ 
\hline
\multirow{2}{*}{Statistical uncertainties} & \multirow{2}{*}{$0.025$} & \multirow{2}{*}{$0.027$} & \multirow{2}{*}{$0.16$}& \multirow{2}{*}{$0.11$}& \multirow{2}{*}{$0.057$}& \multirow{2}{*}{$0.152$} & \multirow{2}{*}{$0.096$} \\ 
           & & & & & & &  \\ 
\hline
Angular acceptance & \multirow{2}{*}{$0.018$} & \multirow{2}{*}{$0.008$} & \multirow{2}{*}{$0.02$} & \multirow{2}{*}{$0.01$} & \multirow{2}{*}{$0.009$} & \multirow{2}{*}{$0.017$} & \multirow{2}{*}{$0.008$} \\
{\scriptsize(simulation statistics)}            & & & & & & &  \\
Angular acceptance        & \multirow{2}{*}{$0.015$} & \multirow{2}{*}{$0.007$} & \multirow{2}{*}{$0.17$}& \multirow{2}{*}{$0.10$}& \multirow{2}{*}{$0.007$}& \multirow{2}{*}{---} & \multirow{2}{*}{$0.015$} \\
{\scriptsize(data--simulation differences)}            & & & & & & &  \\
\multirow{2}{*}{$C_{\rm SP}$ factors}     & \multirow{2}{*}{---} & \multirow{2}{*}{$0.001$} & \multirow{2}{*}{---} & \multirow{2}{*}{---} & \multirow{2}{*}{$0.001$} & \multirow{2}{*}{$0.002$} & \multirow{2}{*}{$0.002$} & \\
                       & & & & & & &  \\
\multirow{2}{*}{\dwave contribution}  & \multirow{2}{*}{$0.004$} & \multirow{2}{*}{$0.003$} & \multirow{2}{*}{---}& \multirow{2}{*}{---}& \multirow{2}{*}{$0.002$}& \multirow{2}{*}{$0.015$} & \multirow{2}{*}{$0.002$} \\                    
                          & & & & & & &  \\
Background                  & \multirow{2}{*}{$0.004$} & \multirow{2}{*}{$0.002$} & \multirow{2}{*}{$0.02$}& \multirow{2}{*}{$0.01$}& \multirow{2}{*}{$0.004$}& \multirow{2}{*}{$^{+	0.012	}_{-	0.004	}$} & \multirow{2}{*}{$0.002$} \\
angular model               & & & & & & &  \\
Mass parameters and   & \multirow{2}{*}{---} & \multirow{2}{*}{---} & \multirow{2}{*}{---}& \multirow{2}{*}{---}& \multirow{2}{*}{$0.001$}& \multirow{2}{*}{$0.001$} & \multirow{2}{*}{---} \\
\Bd contamination     & & & & & & &  \\
Mass--\cosTmu   & \multirow{2}{*}{$	0.007	$} & \multirow{2}{*}{$	0.006$} & \multirow{2}{*}{$	0.07	$}& \multirow{2}{*}{$^{	+0.02	}_{	-0.04	}$}& \multirow{2}{*}{$	0.014$}& \multirow{2}{*}{$^{	+0.009	}_{	-0.012	}$} & \multirow{2}{*}{$	0.016	$} \\
correlations     & & & & & & &  \\
\multirow{2}{*}{Fit bias}  & \multirow{2}{*}{---} & \multirow{2}{*}{$0.001$} & \multirow{2}{*}{$0.01$}& \multirow{2}{*}{$0.07$}& \multirow{2}{*}{$0.003$}& \multirow{2}{*}{$0.002$} & \multirow{2}{*}{$0.005$} \\                    
                          & & & & & & &  \\
Detection & \multirow{2}{*}{---} & \multirow{2}{*}{---} & \multirow{2}{*}{---}& \multirow{2}{*}{---}& \multirow{2}{*}{$0.005$}& \multirow{2}{*}{$0.005$} & \multirow{2}{*}{$0.006$} \\
asymmetry & & & & & & &  \\
Production & \multirow{2}{*}{---} & \multirow{2}{*}{---} & \multirow{2}{*}{---}& \multirow{2}{*}{---}&\multirow{2}{*}{---} & \multirow{2}{*}{---} & \multirow{2}{*}{---} \\
asymmetry & & & & & & &  \\ [+0.04in]
\hline 
Quadratic sum of &   \multirow{2}{*}{$ 0.025 $} & \multirow{2}{*}{$	0.013 $} & \multirow{2}{*}{$0.19$}& \multirow{2}{*}{$^{+	0.012	}_{-	0.013	}$}& \multirow{2}{*}{$0.020$}& \multirow{2}{*}{$^{+	0.028	}_{-	0.027	}$} & \multirow{2}{*}{$ 0.025 $} \\
systematic uncertainties & & & & & & &  \\ \hline
\multirow{2}{*}{Total uncertainties} & \multirow{2}{*}{$0.035$} & \multirow{2}{*}{$0.030$} & \multirow{2}{*}{$0.25$}& \multirow{2}{*}{$^{+	0.016	}_{-	0.017	}$}& \multirow{2}{*}{$0.060$}& \multirow{2}{*}{$0.154$} & \multirow{2}{*}{$0.099$} \\  
  & & & & & & &  \\
\hline
\end{tabular}}
\label{tab:systPwave}
\end{center}
\end{table}
\begin{table}[htbp]
\caption[]{Summary of the measured \BsJpsiKst~\swave properties and their statistical and
systematic uncertainties. 
When no value is given, it means an uncertainty below $5\times10^{-4}$, except for the four phases related to the \swave component, $\delta_{\rm S}$ (rad), in which case the uncertainty is below $5\times10^{-3}$. The \mkpi binning definition is identical to the one given in \tabref{tab:CSPfactors}.}
\begin{center}
\footnotesize
\begin{tabular}{lccccccccc}
\hline 
\noalign{\vskip 0.02 in}
\multirow{2}{*}{} &  \multirow{2}{*}{\ACPS} & \multicolumn{2}{c}{$\mkpi^{\rm bin0}$} & \multicolumn{2}{c}{$\mkpi^{\rm bin1}$} & \multicolumn{2}{c}{$\mkpi^{\rm bin2}$} & \multicolumn{2}{c}{$\mkpi^{\rm bin3}$} \\ 
 & & \multicolumn{1}{c}{$F_{\rm S}$} & \multicolumn{1}{c}{$\delta_{\rm S}$} & $F_{\rm S}$ & \multicolumn{1}{c}{$\delta_{\rm S}$} & $F_{\rm S}$ & \multicolumn{1}{c}{$\delta_{\rm S}$} & $F_{\rm S}$ & $\delta_{\rm S}$ \\ [+0.04in]
\hline
\multirow{2}{*}{Fitted value} & \multirow{2}{*}{$0.167$} & \multirow{2}{*}{$0.475$} & \multirow{2}{*}{$0.54$} & \multirow{2}{*}{$0.080$}& \multirow{2}{*}{$-0.53$}& \multirow{2}{*}{$0.044$}& \multirow{2}{*}{$-1.46$} & \multirow{2}{*}{$0.523$} & \multirow{2}{*}{$-1.76$} \\ 
           & & & & & & & & & \\ 
\hline
\multirow{2}{*}{Statistical uncertainties} & \multirow{2}{*}{$0.114$} & \multirow{2}{*}{$^{+0.108}_{-0.112}$} & \multirow{2}{*}{$0.16$}& \multirow{2}{*}{$^{+0.031}_{-0.025}$}& \multirow{2}{*}{$^{+0.25}_{-0.21}$}& \multirow{2}{*}{$^{+0.042}_{-0.029}$} & \multirow{2}{*}{$^{+0.22}_{-0.19}$} & \multirow{2}{*}{$^{+0.109}_{-0.112}$} & \multirow{2}{*}{$^{+0.13}_{-0.14}$} \\ 
           & & & & & & & & & \\ 
\hline
Angular acceptance & \multirow{2}{*}{$0.028$} & \multirow{2}{*}{$0.039$} & \multirow{2}{*}{$0.03$} & \multirow{2}{*}{$0.012$} & \multirow{2}{*}{$0.065$} & \multirow{2}{*}{$0.015$} & \multirow{2}{*}{$0.10$} & \multirow{2}{*}{$0.065$}& \multirow{2}{*}{$0.06$} \\
{\scriptsize(simulation statistics)}                     & & & & & & & & & \\
Angular acceptance        &  \multirow{2}{*}{$0.015$} & \multirow{2}{*}{$0.058$} & \multirow{2}{*}{$0.08$} & \multirow{2}{*}{$0.019$}& \multirow{2}{*}{$0.18$}& \multirow{2}{*}{$0.027$}& \multirow{2}{*}{$0.27$} & \multirow{2}{*}{$0.006$} & \multirow{2}{*}{$0.04$} \\ 
{\scriptsize(data--simulation differences)}             & & & & & & & & & \\
\multirow{2}{*}{$C_{\rm SP}$ factors}     & \multirow{2}{*}{---} & \multirow{2}{*}{$0.002$} & \multirow{2}{*}{$0.01$} & \multirow{2}{*}{$0.001$} & \multirow{2}{*}{---} & \multirow{2}{*}{$0.002$} & \multirow{2}{*}{---} & \multirow{2}{*}{$0.001$} & \multirow{2}{*}{$0.01$} \\
                        & & & & & & & & & \\
\multirow{2}{*}{\dwave contribution}  & \multirow{2}{*}{$0.008$} & \multirow{2}{*}{$0.010$} & \multirow{2}{*}{$0.02$}& \multirow{2}{*}{$0.005$}& \multirow{2}{*}{$0.03$}& \multirow{2}{*}{$0.008$} & \multirow{2}{*}{$0.08$} & \multirow{2}{*}{$0.002$} & \multirow{2}{*}{$0.04$} \\                    
                           & & & & & & & & & \\
Background                  & \multirow{2}{*}{$0.001$} & \multirow{2}{*}{$	0.002	$} & \multirow{2}{*}{$0.01$}& \multirow{2}{*}{$^{+	0.000	}_{-	0.001	}$}& \multirow{2}{*}{$0.01$}& \multirow{2}{*}{---} & \multirow{2}{*}{$0.03$} & \multirow{2}{*}{$^{+	0.002	}_{-	0.000	}$} & \multirow{2}{*}{$^{+	0.07	}_{-	0.04	}$}\\
angular model                & & & & & & & & & \\
Mass parameters and   & \multirow{2}{*}{$0.001$} & \multirow{2}{*}{$0.001$} & \multirow{2}{*}{$0.01$}& \multirow{2}{*}{---}& \multirow{2}{*}{---} & \multirow{2}{*}{---} & \multirow{2}{*}{---} & \multirow{2}{*}{---} & \multirow{2}{*}{---} \\
\Bd contamination      & & & & & & & & & \\
Mass--\cosTmu   & \multirow{2}{*}{$^{+	0.023	}_{-	0.029	}$} & \multirow{2}{*}{$^{+	0.040	}_{-	0.028	}$} & \multirow{2}{*}{$0.05$}& \multirow{2}{*}{$0.003$}& \multirow{2}{*}{$0.04$}& \multirow{2}{*}{$^{+	0.006	}_{-	0.016	}$} & \multirow{2}{*}{$0.02$} & \multirow{2}{*}{$^{+	0.009	}_{-	0.011	}$} & \multirow{2}{*}{$0.03$}  \\
correlations      & & & & & & & & & \\
\multirow{2}{*}{Fit bias}  & \multirow{2}{*}{$0.004$} & \multirow{2}{*}{$0.005$} & \multirow{2}{*}{$0.01$}& \multirow{2}{*}{$0.003$}& \multirow{2}{*}{$0.02$}& \multirow{2}{*}{$0.007$} & \multirow{2}{*}{$0.032$} & \multirow{2}{*}{$0.015$} & \multirow{2}{*}{$0.01$} \\                    
                           & & & & & & & & & \\
Detection & \multirow{2}{*}{$0.005$} & \multirow{2}{*}{---} & \multirow{2}{*}{---}& \multirow{2}{*}{---}& \multirow{2}{*}{---}& \multirow{2}{*}{---} & \multirow{2}{*}{---} & \multirow{2}{*}{---} & \multirow{2}{*}{---} \\
asymmetry  & & & & & & & & & \\
Production & \multirow{2}{*}{---} & \multirow{2}{*}{---} & \multirow{2}{*}{---}& \multirow{2}{*}{---}& \multirow{2}{*}{---}& \multirow{2}{*}{---} & \multirow{2}{*}{---} & \multirow{2}{*}{---} & \multirow{2}{*}{---} \\
asymmetry & & & & & & & & & \\ [+0.04in]
\hline 
Quadratic sum of & \multirow{2}{*}{$^{+	0.041	}_{-	0.044	}$} & \multirow{2}{*}{$^{+	0.081	}_{-	0.076	}$} & \multirow{2}{*}{$0.10$}& \multirow{2}{*}{$0.023$}& \multirow{2}{*}{$ 0.20 $}& \multirow{2}{*}{$^{+	0.033	}_{-	0.036	}$} & \multirow{2}{*}{$0.30$} & \multirow{2}{*}{$	0.068 $} & \multirow{2}{*}{$^{+	0.11	}_{-	0.09	}$} \\  
systematic uncertainties & & & & & & & & & \\ \hline
\multirow{2}{*}{Total uncertainties} & \multirow{2}{*}{$^{+	0.120	}_{-	0.122	}$} & \multirow{2}{*}{$0.135$} & \multirow{2}{*}{$0.19$}& \multirow{2}{*}{$^{+	0.039	}_{-	0.034	}$}& \multirow{2}{*}{$^{+	0.32	}_{-	0.29	}$}& \multirow{2}{*}{$^{+	0.054	}_{-	0.047	}$} & \multirow{2}{*}{$^{+	0.37	}_{-	0.35	}$} & \multirow{2}{*}{$^{+	0.128	}_{-	0.131	}$} & \multirow{2}{*}{$0.17$} \\  
   & & & & & & & & & \\
\hline
\end{tabular}
\label{tab:systSwave}
\end{center}
\end{table}

\subsubsection{Systematic uncertainties related to the mass fit model}\label{sec:massFitSystematics}

To determine the systematic uncertainty arising from the fixed parameters in the description of the $\jpsi\Km\pip$ invariant mass, these parameters are varied inside their uncertainties, as determined from fits to simulated events.
The fit is then repeated and the widths of the \Bs and \Bd yield distributions are taken as systematic uncertainties on the branching fractions. 
Correlations among the parameters obtained from simulation are taken into account in this procedure. 
For each new fit to the $\jpsi\Km\pip$ invariant mass, the corresponding set of \sweights is calculated and the fit to the weighted angular distributions is repeated.
The widths of the distributions are taken as systematic uncertainties on the angular parameters.
In addition, a systematic uncertainty is added to account for imperfections in the modelling of the upper tail of the \Bd and \Bs peaks. 
Indeed, in the Hypatia distribution model, the parameters $a_2$ and $n_2$ take into account effects such as decays in flight of the hadron, that affect the lineshape of the upper tail and could modify the \Bd leakage into the \Bs peak. 
The estimate of this leakage is recalculated for extreme values of those parameters, and the maximum spread is conservatively added as a systematic uncertainty. 

Systematic uncertainties due to the fixed yields of the \BsJpsiKK, \mbox{\BsJpsipipi}, \BdJpsipipi, and \LbJpsipK peaking backgrounds,\footnote{The yields of the subtracted backgrounds can be considered as fixed, since the sum of negative weights used to subtract them is constant in the nominal fit.} are evaluated by repeating the fit to the invariant mass varying the normalisation of all background sources by either plus or minus one standard deviation of its estimated yield. 
For each of the new mass fits, the angular fit is repeated using the corresponding new sets of \sweights. 
The deviations on each of the angular parameters are then added in quadrature.

Correlations between the $\jpsi K^-\pi^+$ invariant mass and the cosine of the helicity angle \thetamu are taken into account in the nominal fit model, where the mass fit is performed in five bins of \cosTmu.
In order to evaluate systematic uncertainties due to these correlations, the mass fit is repeated with the full range of \cosTmu divided into four or six equal bins. 
For each new mass fit, the angular fit is repeated using the corresponding set of \sweights. 
The deviations from the nominal result for each of the variations are summed quadratically and taken as the systematic uncertainty. 

\subsubsection{Systematic uncertainties related to the angular fit model}

In order to account for systematic uncertainties due to the angular acceptance, two distinct effects are considered, as in Ref.~\cite{LHCb-PAPER-2014-059}.
The first is due to the limited size of the simulation sample used in the acceptance estimation. 
It is estimated by varying the normalisation weights 200 times following a Gaussian distribution within a five standard deviation range 
taking into account their correlations. For each of these sets of normalisation weights, the angular fit is repeated, resulting in a distribution for each fitted parameter. 
The width of the resulting parameter distribution is taken as the systematic uncertainty.
Note that in this procedure, the normalisation weights are varied independently in each \mkpi bin.
The second effect, labelled as data-simulation corrections in the tables, accounts for differences between the data and the simulation, using normalisation weights that are determined assuming the amplitudes measured in Ref.~\cite{LHCb-PAPER-2013-023}.  
The difference with respect to the nominal fit is assigned as a systematic uncertainty.
The uncertainties due to the choice of model for the $C_{\rm SP}$ factors are evaluated as the maximum differences observed in the measured parameters when computing the $C_{\rm SP}$ factors with all of the alternative models, as discussed below.
Instead of the nominal propagator for the \swave, a combination of the $K_{0}^{*}(800)^{0}$ and $K_{0}^{*}(1430)^{0}$ resonances with a non-resonant term using the isobar model is considered, as well as a K-matrix~\cite{Aitchison:1972ay} version. 
A pure phase space term is also used, in order to account for the simplest possible parametrisation.
For the \pwave, the alternative propagators considered are the $K^{*}(892)^{0}$ alone and a combination of this contribution with the $K_{1}^{*}(1410)^{0}$ and the $K_{1}^{*}(1430)^{0}$ using the isobar model.

In order to account for the absence of \dwave terms in the nominal fit model a new fit is performed, including a \dwave component, where the related parameters are fixed to the values measured in the $K^{*}_{2}(1430)^{0}$ region. 
The differences in the measured parameters between the results obtained with and without a \dwave component are taken as the corresponding systematic uncertainty.

The presence of biases in the fit model itself is studied using parametric simulation. For this study, 1000 pseudoexperiments were generated and fitted using the nominal shapes, where the generated parameter values correspond to the ones obtained in the fit to data. 
The difference between the generated value and the mean of the distribution of fitted parameter values are treated as a source of systematic uncertainty. 

Finally, the systematic uncertainties due to the fixed values of the detection and production asymmetries are estimated by varying their values by $\pm 1$ standard deviation and repeating the fit.

\subsection{Branching fraction}\label{sec:systBR}

Several sources of systematic uncertainties on the branching fraction measurements are studied, summarised along with the results in \tabref{tab:syst_normalisation}:
systematic uncertainties due to the external parameter $f_d/f_s$ and due to the branching fraction $\BRof{\phi\to K^+K^-}$;
systematic uncertainties due to the ratio of efficiencies obtained from simulation and due to the angular parameters, propagated into the $\omega$ factors (see \secref{sec:systAng});
and systematic uncertainties affecting the \BsJpsiKst and \BdJpsiKst yields, which are determined from the fit to the $\jpsi K^+\pi^-$ invariant mass and described in \secref{sec:systAng}. 
Finally, a systematic uncertainty due to the \BsJpsiPhi yield determined from the fit to the $\jpsi K^+K^-$ invariant mass distribution, described in \secref{sec:BF_norm_JpsiKK}, is also taken into account, where only the effect due to the modelling of the upper tail of the \Bs peak is considered (see \secref{sec:massFitSystematics}). 
For the computation of the absolute branching fraction \BRof\BsJpsiKst (see \secref{sec:averageBsJpsiKst}), two additional systematic sources are taken into account, the uncertainties in the external parameters \BRof\BdJpsiKst and \BRof\BsJpsiPhi.

\begin{table}[h!]
\begin{center}
\caption{Summary of the measured values for the relative branching fractions and their statistical and systematic uncertainties.}
\begin{tabular}{lcc}
\hline \\ [-2.0ex]
Relative branching fraction & {\large $\frac{\BRof\BsJpsiKst}{\BRof\BdJpsiKst}$} (\%) & {\large $\frac{\BRof\BsJpsiKst}{\BRof\BsJpsiPhi}$} (\%) \\ [1.0ex] \hline 
Nominal value & 2.99  & 4.05  \\\hline
Statistical uncertainties & 0.14  & 0.19  \\\hline
Efficiency ratio & 0.04  & 0.05  \\
Angular correction ($\omega$) & 0.09  & 0.07  \\
Mass model (effect on the yield) & 0.06  & 0.08  \\
$f_d/f_s$ & 0.17  & ---  \\
$\BRof{\phi\to K^+K^-}$ & --- & 0.04 \\
Quadratic sum (excluding $f_d/f_s$) & 0.12  & 0.13  \\ \hline
Total uncertainties & 0.25  & 0.23  \\\hline
\end{tabular}
\label{tab:syst_normalisation}
\end{center}
\end{table}

\section{Penguin pollution in \phis}\label{sec:penguins}
\subsection{Information from $\BsJpsiKst$}

Following the strategy proposed in Refs.~\cite{Fleischer:1999zi, Faller:2008gt, DeBruyn:2014oga}, the measured branching fraction, polarisation fractions and \CP asymmetries can be used to quantify the contributions originating from the penguin topologies in $\BsJpsiKst$.
To that end, the transition amplitude for the $\BsJpsiKst$ decay is written in the general form
\begin{equation}\label{Eq:DecAmp_b2ccd}
A\left(\Bs\to(\jpsi{}\Kstarzb)_i\right) = - \lambda \mathcal{A}_i \left[1 - a_ie^{\text{i}\theta_i}e^{\text{i}\gamma}\right]\:,
\end{equation}
where $\lambda = |V_{us}| = 0.22548^{+0.00068}_{-0.00034}$ \cite{Charles:2015gya} and $i$ labels the different polarisation states.
In the above expression, $\mathcal{A}_i$ is a \CP-conserving hadronic matrix element that represents the tree topology, and $a_i$ parametrises the relative contribution from the penguin topologies.
The \CP-conserving phase difference between the two terms is parametrised by $\theta_i$, whereas their weak phase difference is given by the angle $\gamma$ of the Unitarity Triangle.

Both the branching fraction and the \CP asymmetries depend on the penguin parameters $a_i$ 
and $\theta_i$. The dependence of $A^{\CP}_i$ is given by \cite{Fleischer:1999zi}
\begin{equation}\label{Eq:ACP_penguin}
A^{\CP}_i= -\frac{2a_i\sin\theta_i\sin\gamma}{1-2a_i\cos\theta_i\cos\gamma+a_i^{2}}\:.
\end{equation}
To use the branching fraction information an observable is constructed \cite{Fleischer:1999zi}:
\begin{align}\label{Eq:H_penguin}
H_i & \equiv  \frac{1}{\epsilon} \left|\frac{\mathcal{A}'_i}{\mathcal{A}_i}\right|^2
\frac{\Phi\left(\frac{m_{\jpsi}}{m_{\Bs}}, \frac{m_{\phi}}{m_{\Bs}}\right)}{\Phi\left(\frac{m_{\jpsi}}{m_{\Bs}}, \frac{m_{\Kstarzb}}{m_{\Bs}}\right)}
\frac{\BR{\BsJpsiKst}_{\text{theo}}}{\BR{\BsJpsiPhi}_{\text{theo}}}
\frac{f_i}{f'_i}\:, \\ 
& = \frac{1-2a_i \cos\theta_i \cos\gamma+a_i^{2}}{1+2\epsilon a'_i \cos\theta'_i\cos\gamma +\epsilon^2 a_i^{\prime 2}} \:,\nonumber
\end{align}
where $f_i^{(\prime)}$ represents the polarisation fraction, 
\begin{equation}
\epsilon \equiv \frac{\lambda^2}{1-\lambda^2} = 0.0536 \pm 0.0003\; \mbox{\cite{Charles:2015gya}}\:,
\end{equation}
and $\Phi(x,y) = \sqrt{(1-(x-y)^2)(1-(x+y)^2)}$ is the standard two-body phase-space function.
The primed quantities refer to the $\BsJpsiPhi$ channel, while the non-primed ones refer to $\BsJpsiKst$.
The penguin parameters $a'_i$ and $\theta'_i$ are defined in analogy to $a_i$ and $\theta_i$, and parametrise the transition amplitude of the $\BsJpsiPhi$ decay as
\begin{equation}
A\left(\Bs\to (\jpsi{}\phi)_i\right) = \left(1-\frac{\lambda^2}{2}\right)\mathcal{A}'_i \left[1+\epsilon a'_ie^{\text{i}\theta'_i}e^{\text{i}\gamma}\right]\:.
\end{equation}
Assuming \grpsuthree flavour symmetry, and neglecting contributions from exchange and penguin-annihilation topologies,
\footnote{We follow the decomposition introduced in Ref.~\cite{Gronau:1995hm}.}
which are present in $\BsJpsiPhi$ but have no counterpart in $\BsJpsiKst$, we can identify
\begin{equation}\label{Eq:SU3_atheta}
a'_i = a_i\:,\qquad \theta'_i = \theta_i\:.
\end{equation}
The contributions from the additional decay topologies in $\BsJpsiPhi$ can be probed using the decay $\Bd\to\jpsi{}\phi$ \cite{DeBruyn:2014oga}.
The current upper limit on its branching fraction is $\BR{\Bd\to\jpsi{}\phi} < 1.9\times 10^{-7}$ at 90\% confidence level (C.L.)~\cite{LHCB-PAPER-2013-045}, which implies that the size of these additional contributions is small compared to those associated with the penguin topologies.

The $H_i$ observables are constructed in terms of the theoretical branching fractions defined at zero decay time, which differ from the measured time-integrated branching fractions \cite{Kristof} due to the non-zero decay-width difference $\Delta\Gamma_s$ of the \Bs meson system \cite{HFAG}.
The conversion factor between the two branching fraction definitions \cite{Kristof} is taken to be
\begin{equation}
\frac{\BR{B\to f}_{\text{theo}}}{\BR{B\to f}}
= \frac{1-y_s^2}{1-y_s \eta_i \cos(\phis^{\text{SM}})}\:,
\end{equation}
where $\eta_i$ is the \CP eigenvalue of the final state, and $y_s = \Delta\Gamma_s/2\Gamma_s$. 
Taking values for $\Gamma_s$, $\Delta\Gamma_s$ and $\phis^{\text{SM}}$ from Refs.~\cite{HFAG, Charles:2015gya}, the conversion factor is $1.0608 \pm 0.0045$ $(0.9392 \pm 0.0045)$ for the \CP-even (-odd) states.
For the flavour-specific $\BsJpsiKst$ decay $\eta_i = 0$, resulting in a conversion factor of $0.9963 \pm 0.0006$.
The ratios of hadronic amplitudes $|\mathcal{A}'_i/\mathcal{A}_i|$ are calculated in Ref.~\cite{ThesisKDeBruyn} following the method described in Ref.~\cite{Dighe:1998vk} and using the latest results on form factors from Light Cone QCD Sum Rules (LCSR) \cite{Straub:2015ica}.
This leads to
\begin{align*}
H_0 & = 0.98 \pm 0.07\:\text{(stat)} \pm 0.06\:\text{(syst)} \pm 0.26\:(|\mathcal{A}'_i/\mathcal{A}_i|)\:,\\
H_\parallel & = 0.90 \pm 0.14\:\text{(stat)} \pm 0.08\:\text{(syst)} \pm 0.21\:(|\mathcal{A}'_i/\mathcal{A}_i|)\:,\\
H_\perp & = 1.46 \pm 0.14\:\text{(stat)} \pm 0.11\:\text{(syst)} \pm 0.28\:(|\mathcal{A}'_i/\mathcal{A}_i|)\:.
\end{align*}

\begin{figure}[p]
\center
\includegraphics[height=0.30\textheight]{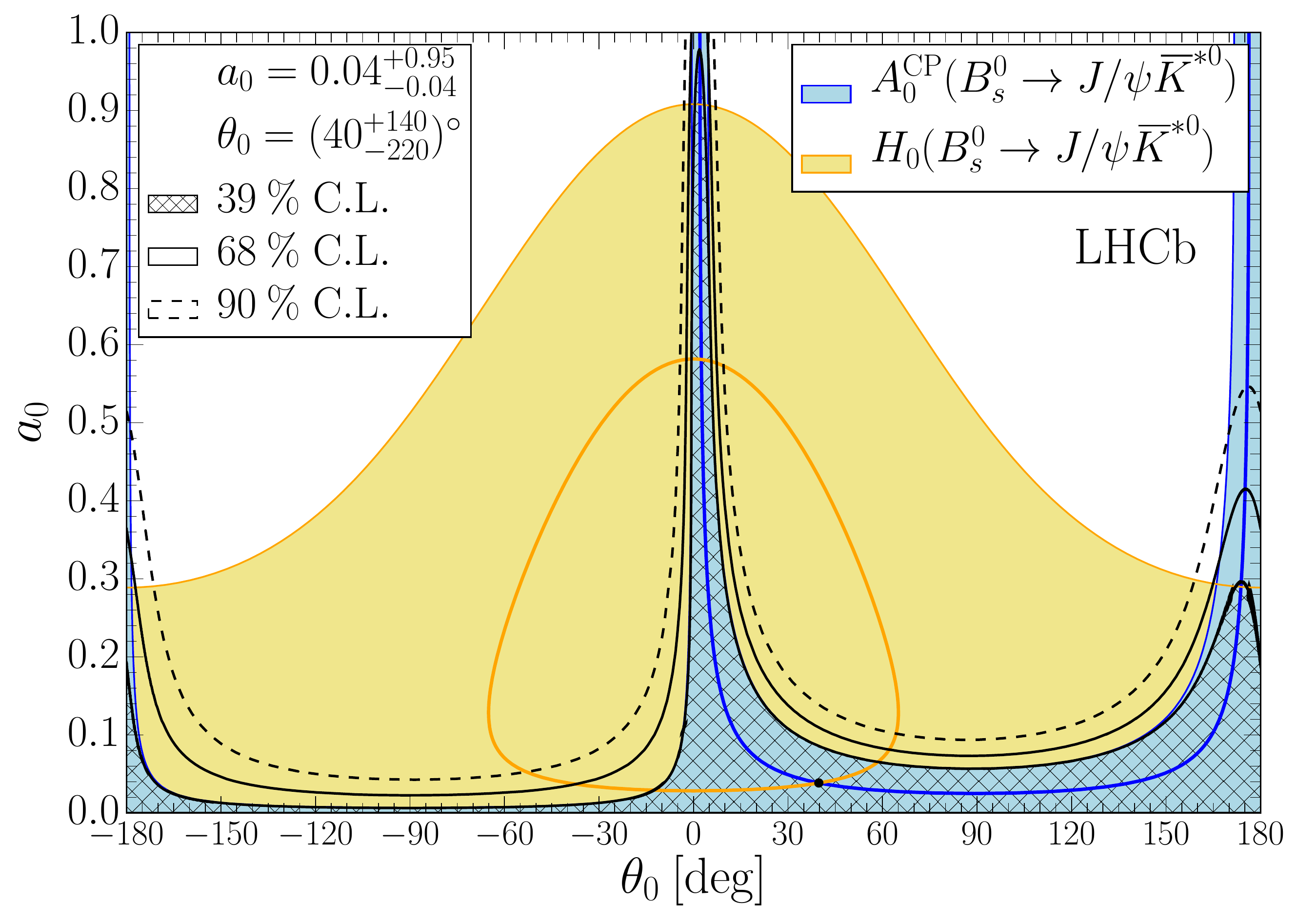}
\includegraphics[height=0.30\textheight]{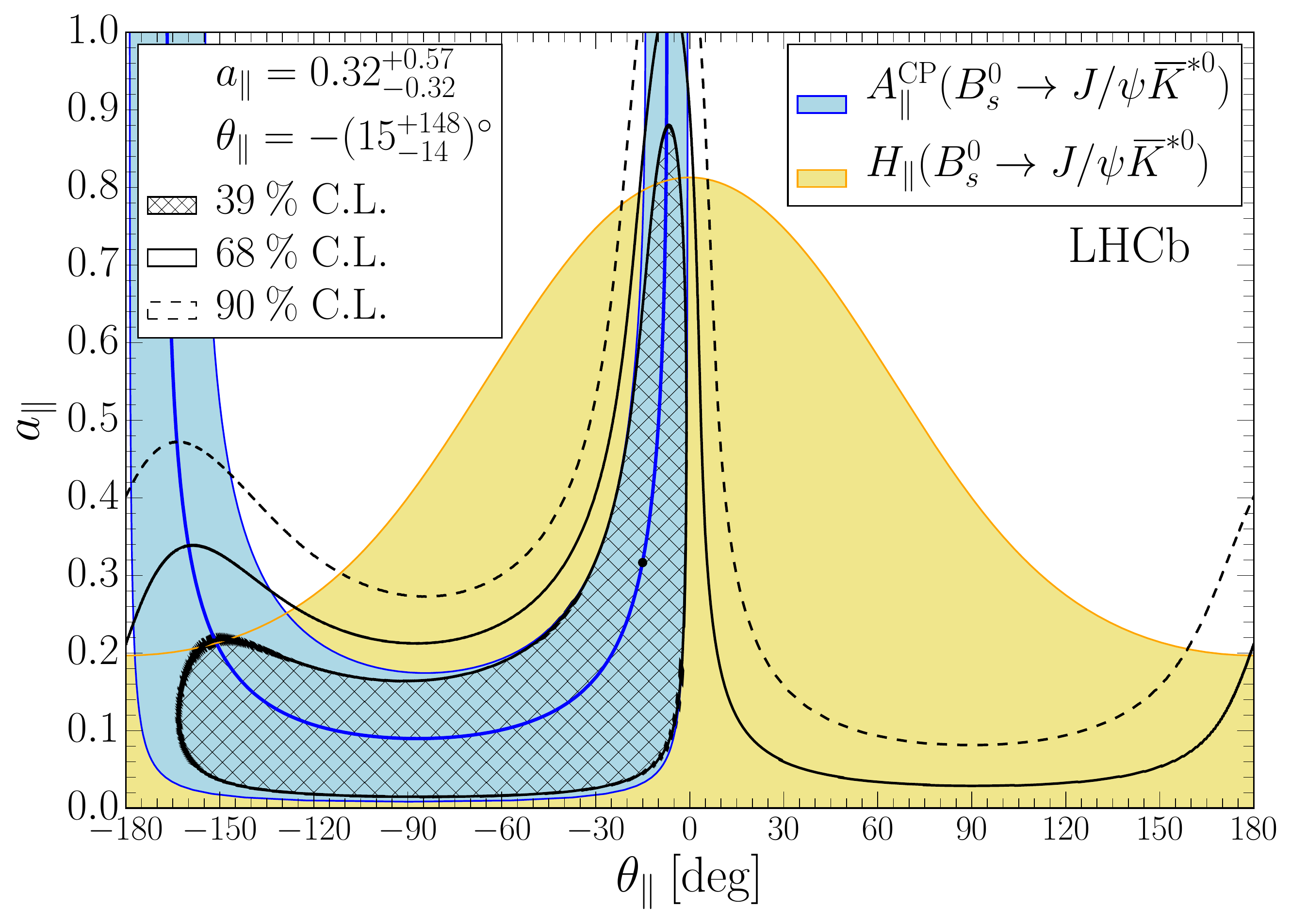}
\includegraphics[height=0.30\textheight]{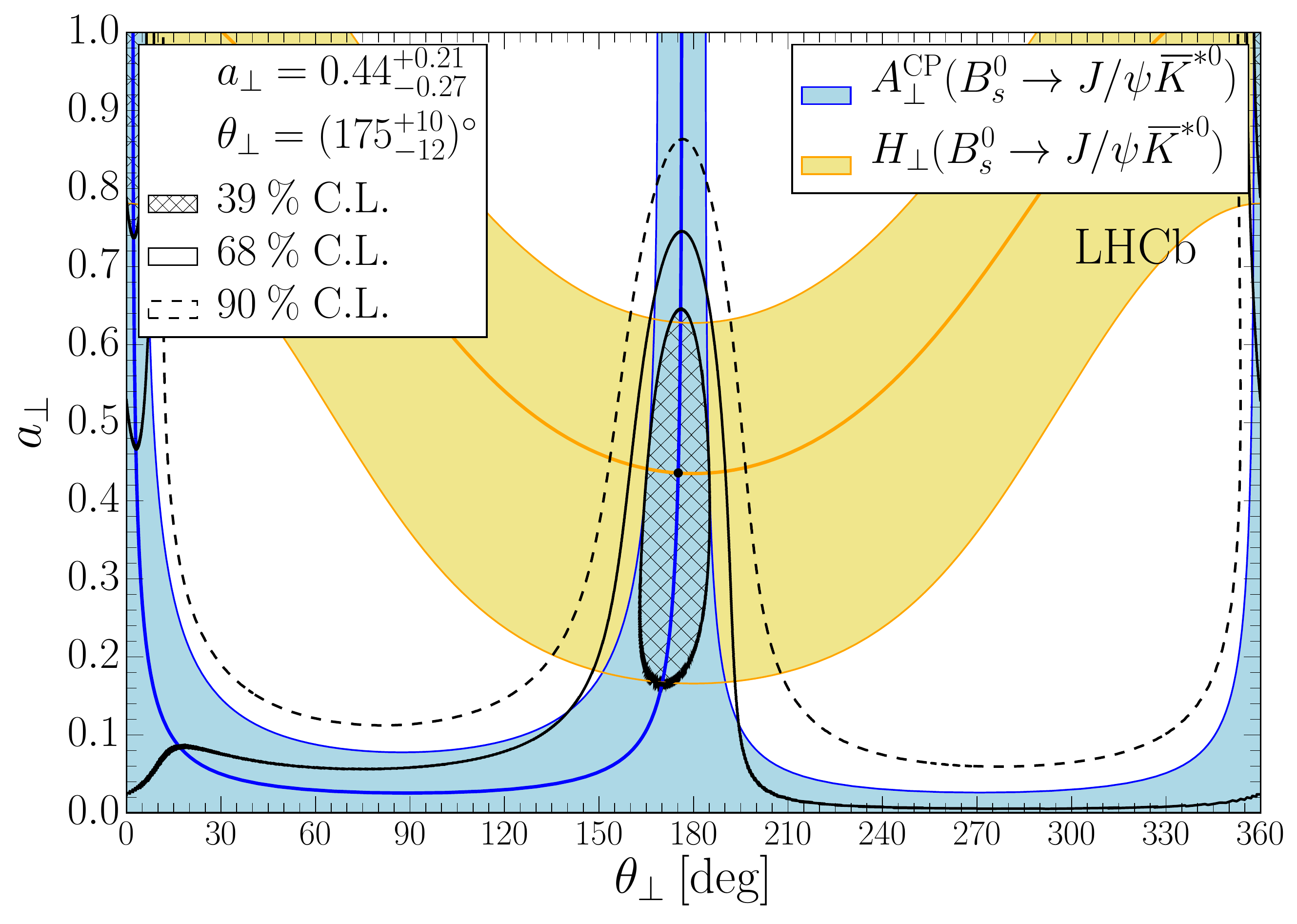}
\caption{Limits on the penguin parameters $a_i$ and $\theta_i$ obtained from intersecting contours derived from the \CP asymmetries and branching fraction information in $\BsJpsiKst$.
Superimposed are the confidence level contours obtained from a $\chi^2$ fit to the data.
Shown are the longitudinal (top), parallel (middle) and perpendicular (bottom) polarisation.}
\label{Fig:PenguinFit_Abs_vs_Ang}
\end{figure}

Assuming \equref{Eq:SU3_atheta} and external input on the Unitarity Triangle angle $\gamma  = \left(73.2_{-7.0}^{+6.3}\right)^{\circ}$ \cite{Charles:2015gya}, the penguin parameters $a_i$ and $\theta_i$ are obtained from a modified least-squares fit to $\{A^{\CP}_i, H_i\}$ in \equref{Eq:ACP_penguin} and \equref{Eq:H_penguin}.
The information on $\gamma$ is included as a Gaussian constraint in the fit.
The values obtained for the penguin parameters are
\begin{align*}
a_0 & = 0.04^{+0.95}_{-0.04}\:, & \theta_0 & = \phantom{-}\left(40^{+140}_{-220}\right)^{\circ}\:,\\
a_\parallel & = 0.32^{+0.57}_{-0.32}\:, & \theta_\parallel & = -\left(15^{+148}_{-14}\right)^{\circ}\:,\\
a_\perp & = 0.44^{+0.21}_{-0.27}\:, & \theta_\perp & = \phantom{-}\left(175^{+11}_{-10}\right)^{\circ}\:.
\end{align*}
For the longitudinal polarisation state the phase $\theta$ is unconstrained.
Correlations between the experimental inputs are ignored, but the effect of including them is small.
The two-dimensional confidence level contours are given in \figref{Fig:PenguinFit_Abs_vs_Ang}.
This figure also shows, as different (coloured) bands, the constraints on the penguin parameters derived from the individual observables entering the $\chi^2$ fit.
The thick inner darker line represents the contour associated with the central value of the input quantity, while the outer darker lines represent the contours associated with the one standard deviation changes.
For the parallel polarisation the central value of the $H$ observable does not lead to physical solutions in the $\theta_\parallel$--$a_\parallel$ plane, and the thick inner line is thus absent.

When decomposed into its different sources, the angle \phis takes the form
\begin{equation}
\phi_{s,i} = -2\beta_s + \phis^{\text{BSM}} + \Delta\phi^{\jpsi{}\phi}_{s,i}(a'_i, \theta'_i)\:,
\end{equation}
where $-2\beta_s$ is the SM contribution, $\phis^{\text{BSM}}$ is a possible BSM phase, and $\Delta\phi^{\jpsi{}\phi}_{s,i}$ is a shift introduced by the presence of penguin pollution in the decay $\BsJpsiPhi$.
In terms of the penguin parameters $a'_i$ and $\theta'_i$, the shift $\Delta\phi^{\jpsi{}\phi}_{s,i}$ is defined as
\begin{equation}\label{Eq:tan_DeltaPhi}
\tan(\Delta\phi^{\jpsi{}\phi}_{s,i}) = \frac{2\epsilon a'_i \cos\theta'_i \sin\gamma+\epsilon^2 a^{\prime 2}_i \sin(2\gamma)}{1+2\epsilon a'_i \cos\theta'_i \cos\gamma+ \epsilon^2 a^{\prime 2}_i \cos(2\gamma)}\:.
\end{equation}
Using Eqs.~\ref{Eq:SU3_atheta} and \ref{Eq:tan_DeltaPhi}, the fit results on $a_i$ and $\theta_i$ given above constrain this phase shift, giving
\begin{alignat*}{4}
\Delta\phi^{\jpsi{}\phi}_{s,0} & = 
\phantom{-}0.003 & ^{+0.084}_{-0.011} & \:\text{(stat)} & ^{+0.014}_{-0.009} & \:\text{(syst)} & ^{+0.047}_{-0.030} &\:(|\mathcal{A}'_i/\mathcal{A}_i|)\:,\\
\Delta\phi^{\jpsi{}\phi}_{s,\parallel} & =
\phantom{-}0.031 & ^{+0.047}_{-0.037} & \:\text{(stat)} & ^{+0.010}_{-0.013} & \:\text{(syst)} & \pm 0.032 & \:(|\mathcal{A}'_i/\mathcal{A}_i|)\:,\\
\Delta\phi^{\jpsi{}\phi}_{s,\perp} & =
-0.045 & \pm 0.012 & \:\text{(stat)} & \pm 0.008 & \:\text{(syst)} & ^{+0.017}_{-0.024} & \:(|\mathcal{A}'_i/\mathcal{A}_i|)\:,
\end{alignat*}
which is in good agreement with the values measured in Ref.~\cite{LHCb-PAPER-2014-058}, and with the
predictions given in Refs.~\cite{DeBruyn:2014oga, Liu:2013nea, Frings:2015eva}.

\begin{figure}[p]
\center
\includegraphics[height=0.30\textheight]{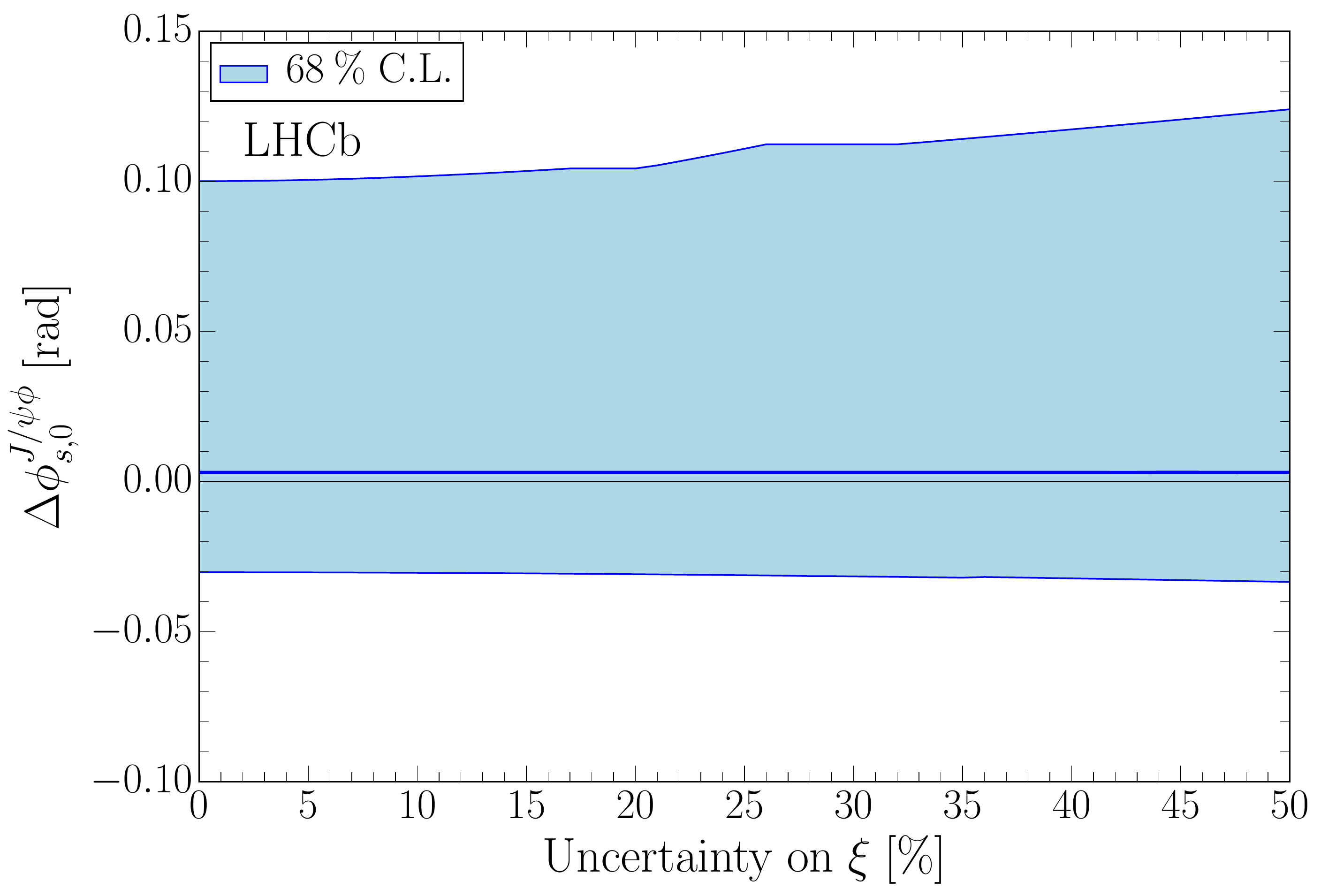}
\includegraphics[height=0.30\textheight]{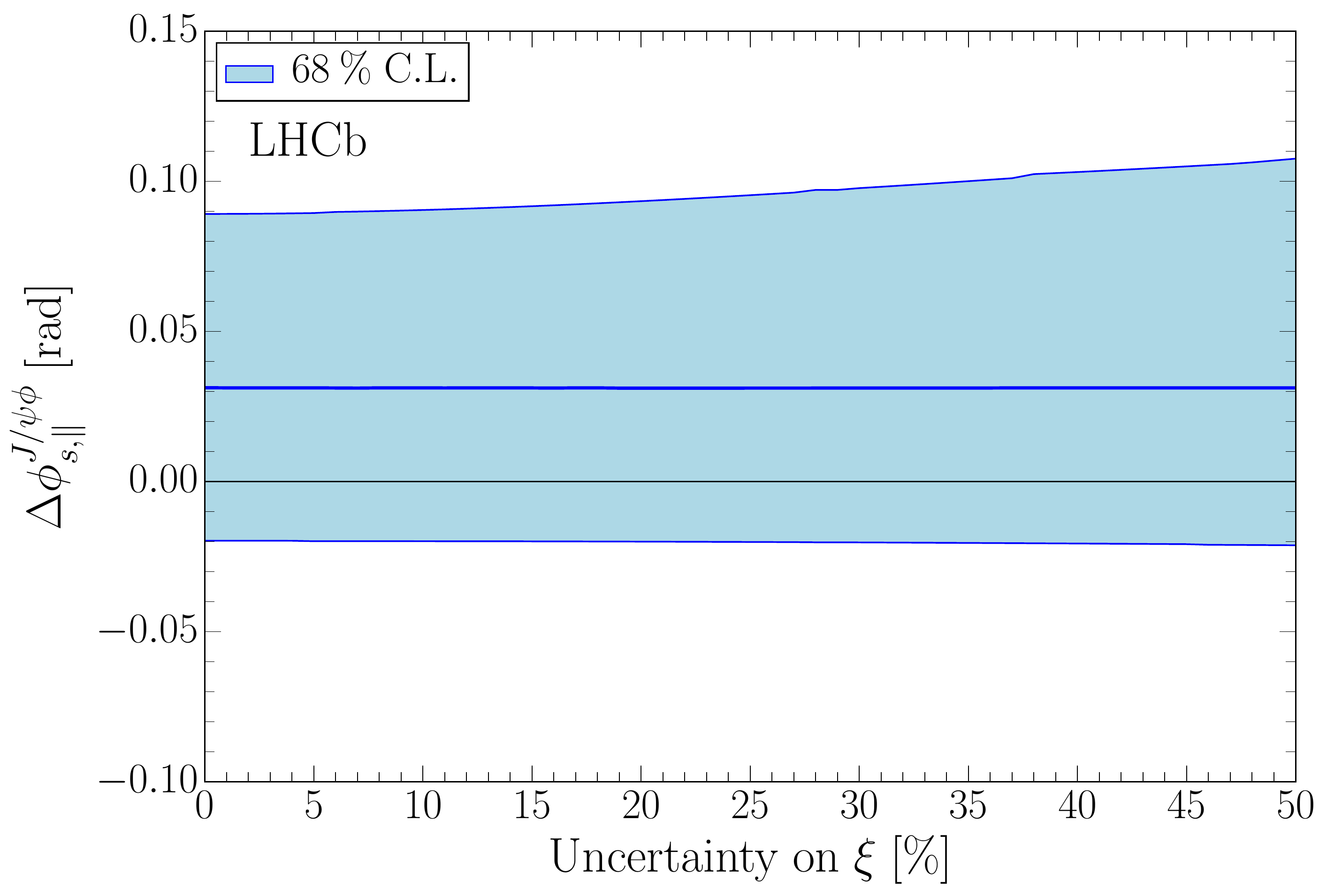}
\includegraphics[height=0.30\textheight]{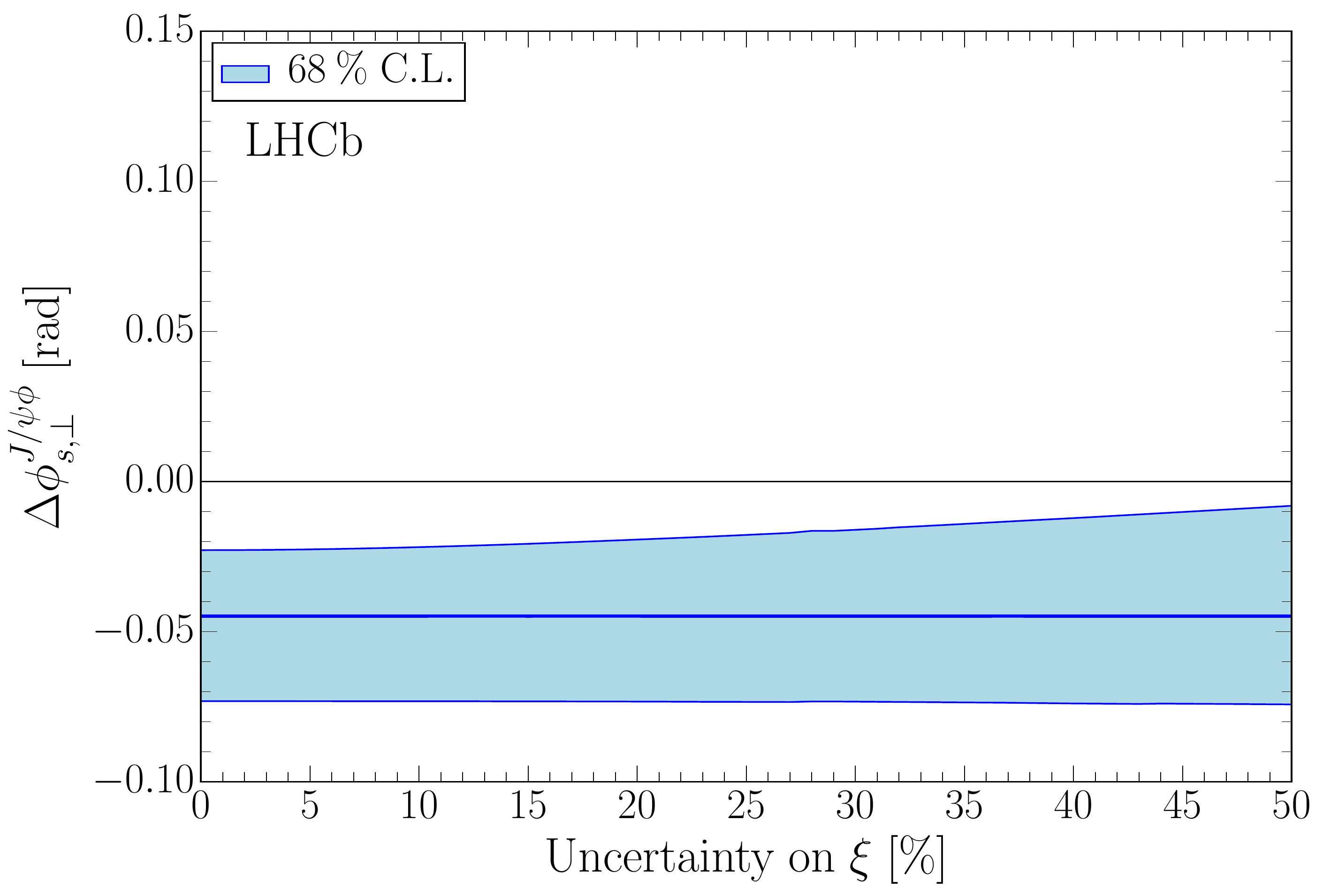}
\caption{Dependence of the uncertainty on the penguin shift $\Delta\phi_{s,i}^{\jpsi\phi}$ on the uncertainty on $\xi$.
The bands correspond to the 68\% C.L.
The longitudinal (top), parallel (middle) and perpendicular (bottom) polarisations are shown.}
\label{Fig:PenguinFit_SU3_Breaking}
\end{figure}

The above results are obtained assuming \grpsuthree flavour symmetry and neglecting contributions from additional decay topologies.
Because $a_i e^{\text{i}\theta_i}$ represents a ratio of hadronic amplitudes, the leading factorisable \grpsuthree-breaking effects cancel, and the relation between $a_i e^{\text{i}\theta_i}$ and $a'_i e^{\text{i}\theta'_i}$ is only affected by non-factorisable \grpsuthree-breaking.
This can be parametrised using two \grpsuthree-breaking parameters $\xi$ and $\delta$ as
\begin{equation}
a'_i = \xi \times a_i\:,\qquad \theta'_i = \theta_i + \delta\:.
\end{equation}
The above quoted results assume $\xi = 1$ and $\delta = 0$.
The dependence of the uncertainty on $\Delta\phi_{s,i}^{\jpsi\phi}$ on the uncertainty on $\xi$ is illustrated in~\figref{Fig:PenguinFit_SU3_Breaking}, while the dependence on the uncertainty on $\delta$ is negligible for the solutions obtained for $\{a_i,\theta_i\}$.

\subsection{Combination with $\Bd\to\jpsi{}\rho^0$}

The information on the penguin parameters obtained from $\BsJpsiKst$ can be complemented with similar information from the \grpsuthree-related mode $\Bd\to\jpsi\rho^0$ \cite{LHCb-PAPER-2014-058}.
Both modes describe a $\bar{b}\to \bar{c}c\bar{d}$ transition, and are related by exchanging the spectator $s\leftrightarrow d$ quarks.
The decay amplitude of $\Bd\to\jpsi\rho^0$ is also parametrised as
\begin{equation}
A\left(\Bd\to(\jpsi{}\rho^0)_i\right) = - \lambda \tilde{\mathcal{A}}_i \left[1 - \tilde{a}_ie^{\text{i}\tilde{\theta}_i}e^{\text{i}\gamma}\right]\:,
\end{equation}
which is the equivalent of~\equref{Eq:DecAmp_b2ccd}.
In contrast to $\BsJpsiKst$, however, $\tilde{a}_i$ and $\tilde{\theta}_i$ also include contributions from exchange and penguin-annihilation topologies, which are present in $\Bd\to\jpsi\rho^0$ but have no counterpart in $\BsJpsiKst$.
Assuming \grpsuthree symmetry, and neglecting the contributions from the additional decay topologies in $\BsJpsiPhi$ and $\Bd\to\jpsi\rho^0$, the relation in \equref{Eq:SU3_atheta} can be extended to
\begin{equation}\label{Eq:SU3_atheta_ext}
a'_i = a_i = \tilde{a}_i\:,\qquad \theta'_i = \theta_i = \tilde{\theta}_i\:,
\end{equation}
which allows a combined fit to be performed to the \CP asymmetries and branching fraction information in $\BsJpsiKst$ and $\Bd\to\jpsi\rho^0$.

The $\Bd\to\jpsi\rho^0$ decay exhibits decay-time-dependent \CP violation, which is described by two parameters, the direct \CP asymmetry $C_i$, which in the \grpsuthree limit is related to $A_i^{\CP}$ as  $C_i = - A_i^{\CP}$, and the mixing-induced \CP asymmetry $S_i$.
Their dependence on the penguin parameters $\tilde{a}_i$ and $\tilde{\theta}_i$ is given by
\begin{align}
C_i & =\phantom{-\eta_f~~}\frac{2\,\tilde{a}_i \sin\tilde{\theta}_i\sin\gamma}{1-2\,\tilde{a}_i\cos\tilde{\theta}_i\cos\gamma+\tilde{a}_i^2}\:,\\
S_i & = -\eta_i\left[\frac{\sin\phi_d-2\,\tilde{a}_i \cos\tilde{\theta}_i\sin(\phi_d+\gamma)+\tilde{a}_i^2\sin(\phi_d+2\gamma)}{1-2\,\tilde{a}_i\cos\tilde{\theta}_i\cos\gamma+\tilde{a}_i^2}\right]\:,
\end{align}
where $\eta_i$ is the polarisation-dependent \CP eigenvalue of the $\Bd\to\jpsi\rho^0$ decay, and $\phi_d$ is a \CP-violating phase arising from the interference between $\Bd$--$\Bdb$ mixing and the subsequent $\Bd$ decay.
The use of $S_i$ to constrain the penguin parameters $a_i$ and $\theta_i$ requires external information on the \CP phase $\phi_d$.
The most precise value of $\phi_d$ is determined from $\Bd\to\jpsi K^0$ decays, but this determination is also affected by penguin pollution.
A recent study of the penguin effects in $\Bu\to\jpsi\pip$, $\Bu\to\jpsi\Kp$, $\Bd\to\jpsi\piz$ and $\Bd\to\jpsi\KS$ decays is performed in Ref.~\cite{DeBruyn:2014oga}, with the latest numerical update \cite{ThesisKDeBruyn}, including the results from Refs.~\cite{LHCb-PAPER-2015-004,LHCb-PAPER-2015-005,Charles:2015gya}, leading to $\phi_d = 0.767 \pm 0.029\rad$.

In addition, a second set of $H_i$ observables can be constructed by replacing \mbox{$\BsJpsiKst$} by $\Bd\to\jpsi\rho^0$ in \equref{Eq:H_penguin}.
To minimise the theoretical uncertainties associated with the use of these $H_i$ observables, the strategy proposed in Ref.~\cite{DeBruyn:2014oga} is adopted.
That is, the relation
\begin{equation}\label{Eq:B2VV_assumption}
\left|\frac{\mathcal{A}'_i}{\mathcal{A}_i}\right| \equiv
\left|\frac{\mathcal{A}'_i(\BsJpsiPhi)}{\mathcal{A}_i(\BsJpsiKst)}\right| = 
\left|\frac{\mathcal{A}'_i(\BsJpsiPhi)}{\mathcal{A}_i(\Bd\to\jpsi\rho^0)}\right|
\end{equation}
between the hadronic amplitudes in $\BsJpsiKst$ and $\Bd\to\jpsi\rho^0$ is assumed, and therefore relying on theoretical input from LCSR is no longer needed.
Instead, the ratio $|\mathcal{A'}/\mathcal{A}|$ can be determined directly from the fit, providing experimental information on this quantity.
Effectively, the three \CP asymmetry parameters $A_i^{\CP}$, $C_i$ and $S_i$ determine the penguin parameters $a_i$ and $\theta_i$.
Thus, this result for $a_i$ and $\theta_i$ predicts the values of the two observables $H_i(\BsJpsiKst)$ and $H_i(\Bd\to\jpsi\rho^0)$.
By comparing these two quantities with the branching fraction and polarisation information on \BsJpsiKst, $\Bd\to\jpsi\rho^0$ and $\BsJpsiPhi$, the hadronic amplitude ratios $|\mathcal{A}'_i/\mathcal{A}_i|$ can be determined.
The impact of the $H_i$ observables on the penguin parameters $a_i$ and $\theta_i$ is negligible in the combined fit.

\begin{figure}[p]
\center
\includegraphics[height=0.30\textheight]{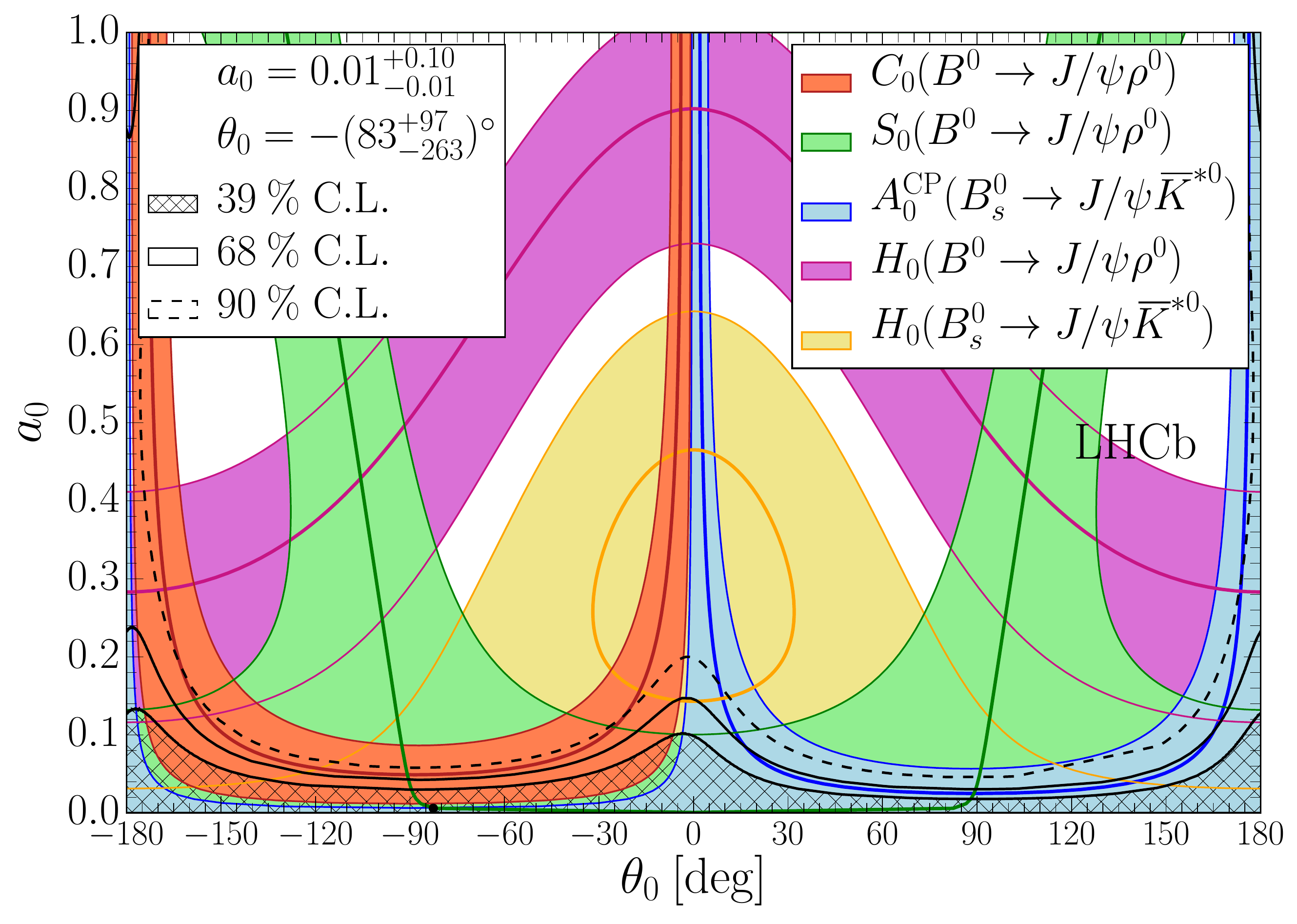}
\includegraphics[height=0.30\textheight]{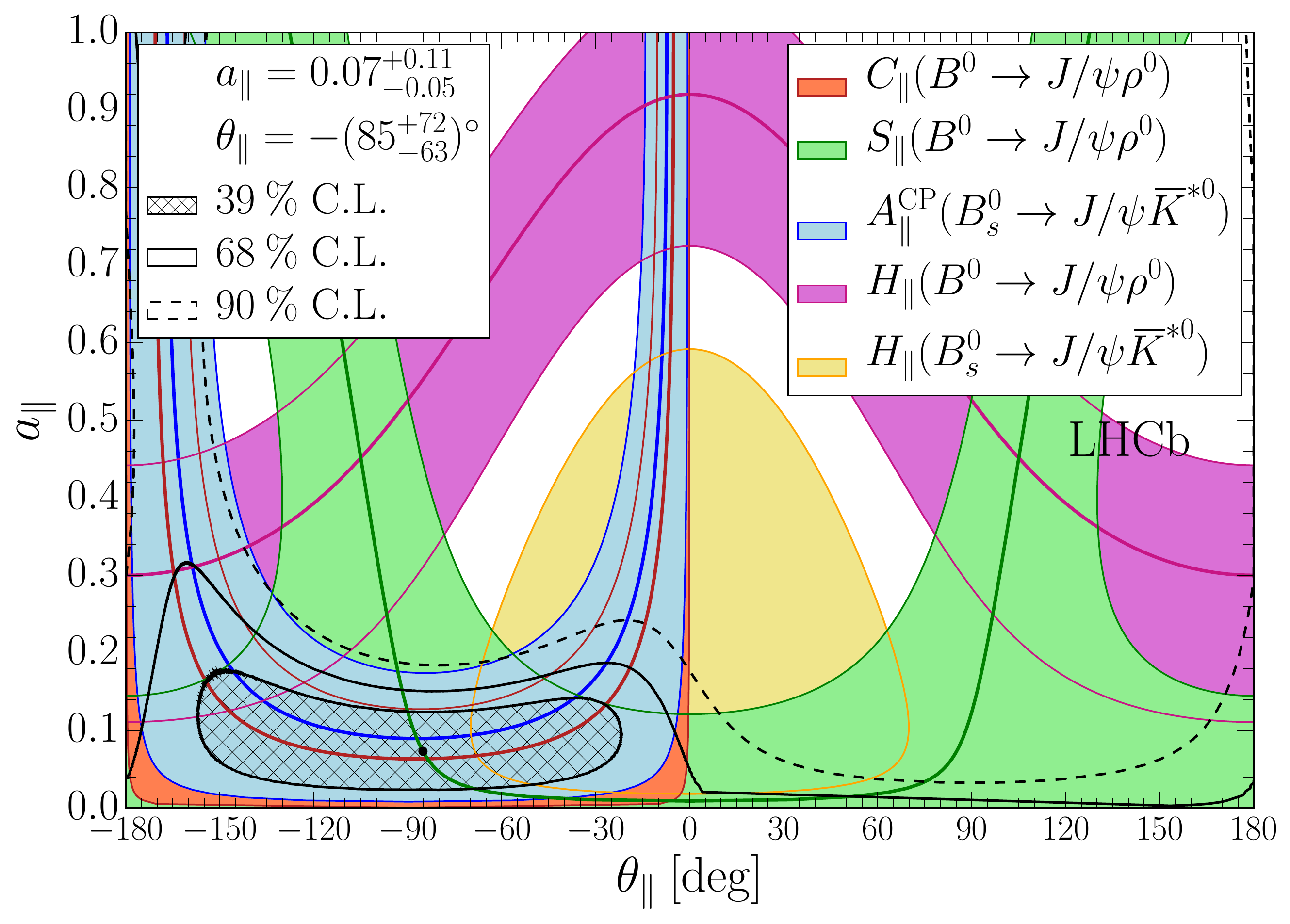}
\includegraphics[height=0.30\textheight]{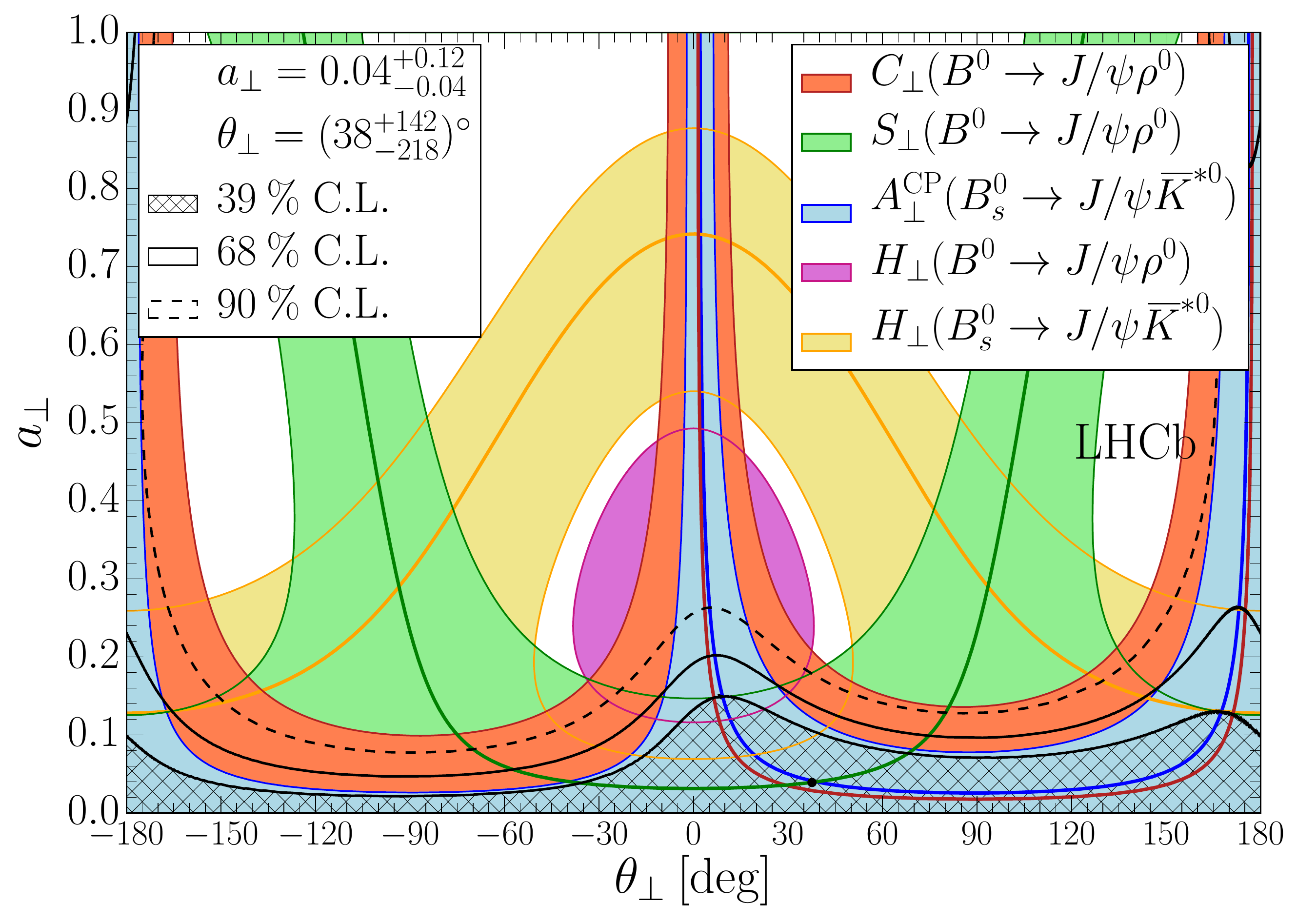}
\caption{Limits on the penguin parameters $a_i$ and $\theta_i$ obtained from intersecting contours derived from the \CP asymmetries and branching fraction information in $\BsJpsiKst$ and $\Bd\to\jpsi\rho^0$.
Superimposed are the confidence level contours obtained from a $\chi^2$ fit to the data.
The longitudinal (top), parallel (middle) and perpendicular (bottom) polarisations are shown.}
\label{Fig:B2VVFit_Abs_vs_Ang}
\end{figure}

For the combined analysis of $\BsJpsiKst$ and $\Bd\to\jpsi\rho^0$ a modified least-squares fit is performed.
External inputs on $\gamma  = \left(73.2_{-7.0}^{+6.3}\right)^{\circ}$ \cite{Charles:2015gya} and $\phi_d = 0.767 \pm 0.029\rad$ \cite{ThesisKDeBruyn} are included as Gaussian constraints in the fit.
The values obtained from the fit are
\begin{align*}
a_0 & = 0.01^{+0.10}_{-0.01}\:, & \theta_0 & = -\left(83^{+97}_{-263}\right)^{\circ}\:,
& \left|\frac{\mathcal{A}'_0}{\mathcal{A}_0}\right| & = 1.195^{+0.074}_{-0.056}\:,\\
a_\parallel & = 0.07^{+0.11}_{-0.05}\:, & \theta_\parallel & = -\left(85^{+72}_{-63}\right)^{\circ}\:,
& \left|\frac{\mathcal{A}'_\parallel}{\mathcal{A}_\parallel}\right| & = 1.238^{+0.104}_{-0.080}\:,\\
a_\perp & = 0.04^{+0.12}_{-0.04}\:, & \theta_\perp & = \phantom{-}\left(38^{+142}_{-218}\right)^{\circ}\:,
& \left|\frac{\mathcal{A}'_\perp}{\mathcal{A}_\perp}\right| & = 1.042^{+0.081}_{-0.063}\:,
\end{align*}
with the two-dimensional confidence level contours given in \figref{Fig:B2VVFit_Abs_vs_Ang}, which also shows the constraints on the penguin parameters derived from the individual observables entering the $\chi^2$ fit as different bands.
Note that the plotted contours for the two $H$ observables do not include the uncertainty due to $|\mathcal{A}'/\mathcal{A}|$.

The results on the penguin phase shift derived from the above results on $a_i$ and $\theta_i$ are
\begin{alignat*}{2}
\Delta\phi^{\jpsi{}\phi}_{s,0} & = 0.000^{+0.009}_{-0.011}\:\text{(stat)} & ^{+0.004}_{-0.009} & \:\text{(syst)}\rad\:,\\
\Delta\phi^{\jpsi{}\phi}_{s,\parallel} & = 0.001^{+0.010}_{-0.014}\:\text{(stat)} & \:\pm 0.008 & \:\text{(syst)}\rad\:,\\
\Delta\phi^{\jpsi{}\phi}_{s,\perp} & = 0.003^{+0.010}_{-0.014}\:\text{(stat)} & \:\pm 0.008 & \:\text{(syst)}\rad\:.
\end{alignat*}
These results are dominated by the input from the \CP asymmetries in $\Bd\to\jpsi\rho^0$, and show that the penguin pollution in the determination of $\phis$ is small.

\section{Conclusions}
\label{sec:conclusion}

Using the full \lhcb Run I data sample, the branching fraction, the polarisation fractions and the direct \CP violation parameters in \BsJKst decays have been measured. 
The results are
\begin{equation*}
\BR{\BsJpsiKst} = \left(4.14 \pm 0.18 \text{(stat)} \pm  0.26 \text{(syst)} \pm 0.24 (f_d/f_s) \right)\times 10^{-5}
\end{equation*}
  \[
  \setlength{\arraycolsep}{1mm}
  \begin{array}{rclllllll}
    f_0                  &\;=\; & \phantom{-}0.497 & \pm & 0.025 & \text{(stat)} & \pm & 0.025 & \text{(syst)}  \\
    f_\|                 &\;=\; & \phantom{-}0.179 & \pm & 0.027 & \text{(stat)} & \pm & 0.013 & \text{(syst)}    \\
    A^{\CP}_0(\BsJKst)    &\;=\; & -0.048           & \pm & 0.057 & \text{(stat)} & \pm & 0.020 & \text{(syst)}    \\
    A^{\CP}_\|(\BsJKst)   &\;=\; & \phantom{-}0.171 & \pm & 0.152 & \text{(stat)} & \pm & 0.028 & \text{(syst)}    \\
    A^{\CP}_\perp(\BsJKst) &\;=\; & -0.049           & \pm & 0.096 & \text{(stat)} & \pm & 0.025 & \text{(syst)\:,}    \\
  \end{array}
  \]
which supersede those of Ref.~\cite{LHCb-PAPER-2012-014}, with precision improved by a factor of $2-3$. 
The shift on \phis\ due to penguin pollution is estimated from a combination with the $\Bd\to\jpsi\rho^0$ channel~\cite{LHCb-PAPER-2014-058}, and is found be to compatible with the result from the earlier analysis.

\section*{Acknowledgements}


\noindent We express our gratitude to our colleagues in the CERN
accelerator departments for the excellent performance of the LHC. We
thank the technical and administrative staff at the LHCb
institutes. We acknowledge support from CERN and from the national
agencies: CAPES, CNPq, FAPERJ and FINEP (Brazil); NSFC (China);
CNRS/IN2P3 (France); BMBF, DFG, HGF and MPG (Germany); INFN (Italy); 
FOM and NWO (The Netherlands); MNiSW and NCN (Poland); MEN/IFA (Romania); 
MinES and FANO (Russia); MinECo (Spain); SNSF and SER (Switzerland); 
NASU (Ukraine); STFC (United Kingdom); NSF (USA).
The Tier1 computing centres are supported by IN2P3 (France), KIT and BMBF 
(Germany), INFN (Italy), NWO and SURF (The Netherlands), PIC (Spain), GridPP 
(United Kingdom).
We are indebted to the communities behind the multiple open 
source software packages on which we depend. We are also thankful for the 
computing resources and the access to software R\&D tools provided by Yandex LLC (Russia).
Individual groups or members have received support from 
EPLANET, Marie Sk\l{}odowska-Curie Actions and ERC (European Union), 
Conseil g\'{e}n\'{e}ral de Haute-Savoie, Labex ENIGMASS and OCEVU, 
R\'{e}gion Auvergne (France), RFBR (Russia), XuntaGal and GENCAT (Spain), Royal Society and Royal
Commission for the Exhibition of 1851 (United Kingdom).

\clearpage
{\noindent\normalfont\bfseries\Large Appendices}
\appendix

\section{Angular acceptance}
\label{sec:angularAcc}

To take into account angular acceptance effects, ten normalisation weights, $\xi_{ij}$, are computed and embedded in the normalization integral of the angular distribution given in Eq.~\ref{eq:angPdf} following the procedure described in Ref.~\cite{TristansThesis}. Using the transversity amplitude basis, the fitting \pdf can be written as
\begin{equation}
\label{eq:angNormPdf}
\frac{\text{d}\Gamma(\theta_K,\thetamu,\phihel)}{\text{d}\Omega} = \frac{\sum_{i}\sum_{j \leq i}{\mathcal{R}e{[A_{i}A_{j}^{*}\:F_{ij}(\theta_K,\thetamu,\phihel)]}}}{\sum_{k}\sum_{l \leq k}{\mathcal{R}e{[A_{k} A_{l}^{*}\: \int{F_{kl}(\theta_K',\thetamu',\phihel')\:\epsilon_{\Omega}(\theta_K',\thetamu',\phihel')\:\text{d}\Omega'}}]}},
\end{equation}
where the real or imaginary angular functions $F_{ij}(\theta_K,\thetamu,\phihel)$ are obtained when combining \equref{eq:angPdf} and Eq.~\ref{eq:helAmp1}-\ref{eq:helAmp4}, and where $\epsilon_{\Omega}(\theta_K,\thetamu,\phihel)$ denotes the angular acceptance. The normalization weights correspond to the integrals
\begin{equation}
\label{eq:normWeights}
\xi_{ij} = \\
\begin{cases}
\int{\mathcal{R}e[F_{ij}(\theta_K,\thetamu,\phihel)]\:\epsilon_{\Omega}(\theta_K,\thetamu,\phihel)\:\text{d}\Omega},\; \text{if } F_{ij}\in\mathbb{R}  \,,\\
\int{\mathcal{I}m[F_{ij}(\theta_K,\thetamu,\phihel)]\:\epsilon_{\Omega}(\theta_K,\thetamu,\phihel)\:\text{d}\Omega},\; \text{if } F_{ij}\in\mathbb{I}  \,.\\
\end{cases}
\end{equation}
In the absence of acceptance effects, the normalisation weights related to the interference terms are equal to zero by definition, 
whereas those related to each polarisation amplitude squared are equal to unity.
Eight sets of normalisation weights are calculated separately, one for each \mkpi bin and kaon charge.

\begin{table}[b!]
  \center
  \caption{\small Corrected angular acceptance weights for $K^-\pi^+$ events lying in the first \mkpi bin.
                  The $\xi_{ij}$ weights are normalised with respect to the $\xi_{00}$ weight. }
 
  \begin{tabular}{r l c}
    \hline
    \multicolumn{2}{c}{$ij$}        &  $\xi_{ij}/\xi_{00}$         \\
    \hline
     1   &  (00)                    & $1.000             $\\
     2   &  ($\parallel\parallel$)  & $+1.379 \pm  0.029$   \\
     3   &  ($\perp\perp$)          & $+1.388 \pm  0.030$   \\
     4   &  ($\parallel\perp$)      & $+0.035 \pm  0.019$   \\
     5   &  (0$\parallel$)          & $-0.003 \pm  0.012$   \\
     6   &  (0$\perp$)              & $+0.010 \pm  0.011$   \\
     7   &  (SS)                    & $+1.190 \pm  0.019$   \\
     8   &  (S$\parallel$)          & $-0.042 \pm  0.017$   \\
     9   &  (S$\perp$)              & $+0.029 \pm  0.016$   \\
     10  &  (S0)                    & $-0.929 \pm  0.024$   \\
    \hline
  \end{tabular}
  \label{tab:angAccWeightsUncorr}
\end{table}
In order to correct both for imperfections in the detector simulation and for the absence of any \swave component in the
simulation sample, the weights are refined using an iterative procedure where the angular acceptance is re-evaluated
recursively until it does not change significantly. 
\tabref{tab:angAccWeightsUncorr} gives one set of normalisation weights after the iterative procedure.
The effect of this correction is below one standard deviation for all the normalisation weights except for the (S0) weight.
This is expected due to the rapid efficiency drop close to $\cos \thetaK = 1$ which directly impacts the (S0) weight.
At each step of this procedure
the simulation sample is corrected both for the absence of an \swave component and for the imperfections in the detector simulation.
For the first correction, the angular fit result to data is used, whereas for the second the kaon and muon track momentum distributions of data are used.
In both cases the correction is implemented by assigning weights to each event of the simulation sample.


\section{Correlation matrix}\label{App:CorrMatrix}

The statistical-only correlation matrix of the angular parameters obtained from the fit to data, as described in \secref{sec:systAng}, is given in \tabref{tab:correlationMatrix}.
Here, the superscript $l=0,1,2,3$ in $F_{\rm S}^l$ and $\delta_{\rm S}^l$ represent the number of the \mkpi bin as defined in \tabref{tab:CSPfactors}. 
\begin{table}[b!]
\center
\caption{\label{tab:correlationMatrix} Statistical correlation matrix for the parameters from the angular fit.}
\rotatebox{90}{
\resizebox{0.95\textheight}{!}{
\begin{tabular}{l c c c c c c c c c c c c c c c c}
\hline
\noalign{\vskip 0.02 in}
     &   $\ACPL$     & $\ACPS$     & $\ACPpa$     & $\ACPpe$     & $F_{\rm S}^0$     & $F_{\rm S}^1$     & $F_{\rm S}^2$     & $F_{\rm S}^3$     & $\delpar$     & $\delperp$     & $\delta_{\rm S}^0$     & $\delta_{\rm S}^1$     & $\delta_{\rm S}^2$     & $\delta_{\rm S}^3$     & $\fL$     & $\fpa$    \\
\noalign{\vskip 0.02 in}
\hline
\noalign{\vskip 0.02 in}
$\ACPL$  & $+1.00$      & $-0.12$      & $-0.11$      & $-0.17$      & $-0.13$      & $-0.02$      & $-0.06$      & $-0.01$      & $+0.03$      & $+0.02$      & $+0.10$      & $-0.00$      & $+0.07$      & $+0.01$      & $+0.06$      & $-0.05$    \\
\noalign{\vskip 0.02 in}
$\ACPS$  &           & $+1.00$      & $-0.14$      & $-0.12$      & $+0.16$      & $-0.12$      & $+0.03$      & $-0.10$      & $+0.00$      & $-0.06$      & $+0.02$      & $+0.07$      & $+0.05$      & $+0.07$      & $+0.01$      & $+0.03$    \\
\noalign{\vskip 0.02 in}
$\ACPpa$  &           &           & $+1.00$      & $-0.49$      & $+0.02$      & $+0.09$      & $-0.02$      & $+0.08$      & $+0.09$      & $+0.06$      & $-0.06$      & $-0.04$      & $-0.05$      & $-0.12$      & $-0.04$      & $-0.07$    \\
\noalign{\vskip 0.02 in}
$\ACPpe$  &           &           &           & $+1.00$      & $-0.00$      & $-0.01$      & $-0.06$      & $-0.07$      & $-0.09$      & $-0.03$      & $-0.03$      & $+0.01$      & $-0.02$      & $+0.07$      & $+0.01$      & $-0.06$    \\
\noalign{\vskip 0.02 in}
$F_{\rm S}^0$  &           &           &           &           & $+1.00$      & $+0.01$      & $-0.01$      & $-0.03$      & $-0.10$      & $-0.24$      & $-0.77$      & $+0.01$      & $+0.04$      & $-0.00$      & $+0.10$      & $-0.09$    \\
\noalign{\vskip 0.02 in}
$F_{\rm S}^1$  &           &           &           &           &           & $+1.00$      & $-0.01$      & $-0.00$      & $-0.02$      & $-0.05$      & $-0.01$      & $-0.25$      & $+0.03$      & $-0.01$      & $+0.15$      & $-0.10$    \\
\noalign{\vskip 0.02 in}
$F_{\rm S}^2$  &           &           &           &           &           &           & $+1.00$      & $+0.01$      & $-0.04$      & $+0.07$      & $+0.01$      & $-0.00$      & $-0.22$      & $+0.00$      & $-0.02$      & $+0.04$    \\
\noalign{\vskip 0.02 in}
$F_{\rm S}^3$  &           &           &           &           &           &           &           & $+1.00$      & $+0.08$      & $+0.08$      & $+0.00$      & $-0.01$      & $-0.03$      & $-0.29$      & $-0.09$      & $+0.04$    \\
\noalign{\vskip 0.02 in}
$\delpar$  &           &           &           &           &           &           &           &           & $+1.00$      & $+0.62$      & $+0.10$      & $+0.14$      & $+0.03$      & $+0.11$      & $+0.04$      & $-0.03$    \\
\noalign{\vskip 0.02 in}
$\delperp$  &           &           &           &           &           &           &           &           &           & $+1.00$      & $+0.17$      & $+0.13$      & $-0.02$      & $+0.13$      & $+0.05$      & $-0.04$    \\
\noalign{\vskip 0.02 in}
$\delta_{\rm S}^0$  &           &           &           &           &           &           &           &           &           &           & $+1.00$      & $+0.04$      & $+0.03$      & $+0.04$      & $+0.08$      & $+0.04$    \\
\noalign{\vskip 0.02 in}
$\delta_{\rm S}^1$  &           &           &           &           &           &           &           &           &           &           &           & $+1.00$      & $+0.04$      & $+0.04$      & $+0.13$      & $-0.05$    \\
\noalign{\vskip 0.02 in}
$\delta_{\rm S}^2$  &           &           &           &           &           &           &           &           &           &           &           &           & $+1.00$      & $+0.04$      & $+0.27$      & $-0.08$    \\
\noalign{\vskip 0.02 in}
$\delta_{\rm S}^3$  &           &           &           &           &           &           &           &           &           &           &           &           &           & $+1.00$      & $+0.11$      & $+0.00$    \\
\noalign{\vskip 0.02 in}
$\fL$  &           &           &           &           &           &           &           &           &           &           &           &           &           &           & $+1.00$      & $-0.34$    \\
\noalign{\vskip 0.02 in}
$\fpa$  &           &           &           &           &           &           &           &           &           &           &           &           &           &           &           & $+1.00$    \\
\noalign{\vskip 0.02 in}
\hline
\end{tabular}
}
}
\end{table}
\clearpage
\addcontentsline{toc}{section}{References}
\setboolean{inbibliography}{true}
\bibliographystyle{LHCb}
\bibliography{main,LHCb-PAPER,LHCb-DP.bib}

\newpage
\centerline{\large\bf LHCb collaboration}
\begin{flushleft}
\small
R.~Aaij$^{38}$, 
B.~Adeva$^{37}$, 
M.~Adinolfi$^{46}$, 
A.~Affolder$^{52}$, 
Z.~Ajaltouni$^{5}$, 
S.~Akar$^{6}$, 
J.~Albrecht$^{9}$, 
F.~Alessio$^{38}$, 
M.~Alexander$^{51}$, 
S.~Ali$^{41}$
% ...

{\footnotesize \it
% ...
$ ^{5}$Clermont Universit\'{e}, Universit\'{e} Blaise Pascal, CNRS/IN2P3, LPC, Clermont-Ferrand, France\\
$ ^{6}$CPPM, Aix-Marseille Universit\'{e}, CNRS/IN2P3, Marseille, France\\
% ...
$ ^{9}$Fakult\"{a}t Physik, Technische Universit\"{a}t Dortmund, Dortmund, Germany\\
% ...
$ ^{37}$Universidad de Santiago de Compostela, Santiago de Compostela, Spain\\
$ ^{38}$European Organization for Nuclear Research (CERN), Geneva, Switzerland\\
% ...
$ ^{41}$Nikhef National Institute for Subatomic Physics, Amsterdam, The Netherlands\\
% ...
$ ^{46}$H.H. Wills Physics Laboratory, University of Bristol, Bristol, United Kingdom\\
% ...
$ ^{51}$School of Physics and Astronomy, University of Glasgow, Glasgow, United Kingdom\\
$ ^{52}$Oliver Lodge Laboratory, University of Liverpool, Liverpool, United Kingdom\\
% ...
}
\end{flushleft}

\newpage

\end{document}